\documentclass[fleqn,usenatbib]{mnras}

\usepackage{newtxtext,newtxmath}

\usepackage[T1]{fontenc}

\DeclareRobustCommand{\VAN}[3]{#2}
\let\VANthebibliography\thebibliography
\def\thebibliography{\DeclareRobustCommand{\VAN}[3]{##3}\VANthebibliography}

\usepackage{graphicx}	
\usepackage{amsmath}	
\usepackage{siunitx}
\usepackage{float}
\usepackage{makecell}
\usepackage{soul}

\newcommand{\chandra}{\textit{Chandra}\ }
\DeclareSIUnit\erg{erg}

\title[\chandra and VLA observations of NGC~6652B]{The MAVERIC Survey: Simultaneous \chandra and VLA observations of the transitional millisecond pulsar candidate NGC~6652B}

\author[Alessandro Paduano et al.]{Alessandro Paduano$^{1}$\thanks{E-mail: a.paduano@postgrad.curtin.edu.au},
Arash Bahramian$^{1}$,
James C. A. Miller-Jones$^{1}$,
Adela Kawka$^{1}$,
\newauthor Jay Strader$^{2}$,
Laura Chomiuk$^{2}$,
Craig O. Heinke$^{3}$,
Thomas J. Maccarone$^{4}$,
\newauthor Christopher T. Britt$^{5}$,
Richard M. Plotkin$^{6}$,
Aarran W. Shaw$^{6}$,
Laura Shishkovsky$^{2}$,
\newauthor Evangelia Tremou$^{7}$,
Vlad Tudor$^{1}$,
and Gregory R. Sivakoff$^{3}$
\\
$^{1}$International Centre for Radio Astronomy Research - Curtin University, GPO Box U1987, Perth, WA 6845, Australia\\
$^{2}$Center for Data Intensive and Time Domain Astronomy, Department of Physics and Astronomy, Michigan State University, East Lansing, MI 48824, USA\\
$^{3}$Department of Physics, University of Alberta, CCIS 4-181, Edmonton, AB T6G 2E1, Canada\\
$^{4}$Department of Physics \& Astronomy, Texas Tech University, Box 41051, Lubbock, TX 79409-1051, USA\\
$^{5}$Space Telescope Science Institute, 3700 San Martin Dr, Baltimore, MD 21218, USA\\
$^{6}$Department of Physics, University of Nevada, Reno, NV 89557, USA\\
$^{7}$LESIA, Observatoire de Paris, CNRS, PSL, SU/UPD, Meudon, France
}

\date{Accepted XXX. Received YYY; in original form ZZZ}

\pubyear{2021}

\begin{document}
\label{firstpage}
\pagerange{\pageref{firstpage}--\pageref{lastpage}}
\maketitle

\begin{abstract}
    Transitional millisecond pulsars are millisecond pulsars that switch between a rotation-powered millisecond pulsar state and an accretion-powered X-ray binary state, and are thought to be an evolutionary stage between neutron star low-mass X-ray binaries and millisecond pulsars. So far, only three confirmed systems have been identified in addition to a handful of candidates. We present the results of a multi-wavelength study of the low-mass X-ray binary NGC~6652B in the globular cluster NGC 6652, including simultaneous radio and X-ray observations taken by the Karl G. Jansky Very Large Array and the \chandra X-ray Observatory, and optical spectroscopy and photometry. This source is the second brightest X-ray source in NGC 6652 ($L_{\textrm{X}}\sim$\SI{1.8e34}{\erg\per\second}) and is known to be variable. We observe several X-ray flares over the duration of our X-ray observations, in addition to persistent radio emission and occasional radio flares. Simultaneous radio and X-ray data show no clear evidence of anti-correlated variability. Optical spectra of NGC~6652B indicate variable, broad H $\alpha$ emission which transitions from double-peaked emission to absorption over a time-scale of hours. We consider a variety of possible explanations for the source behaviour, and conclude that based on the radio and X-ray luminosities, short time-scale variability and X-ray flaring, and optical spectra, NGC~6652B is best explained as a transitional millisecond pulsar candidate that displays prolonged X-ray flaring behaviour. However, this could only be confirmed with observations of a change to the rotation-powered millisecond pulsar state.
\end{abstract}




\section{Introduction}
\subsection{Transitional millisecond pulsars}
    Low-mass X-ray binaries (LMXBs) are binary systems containing a compact object, black hole (BH) or neutron star (NS), accreting from a low mass ($M_c<1 \textrm{M}_{\odot}$) companion star. NS LMXBs are responsible for the creation of millisecond pulsars (MSPs), old, ``dead'' radio pulsars recycled through accretion from a companion star \citep{Alpar1982}. In most cases the resultant system ends up with a spin period of a few milliseconds, and such systems form some of the most stable clocks in the Universe.
    
    This link between LMXBs and MSPs was indicated by the detections of accreting millisecond X-ray pulsars, in particular SAX J1808.4-3658 \citep{Wijnands1998a}, before being confirmed by the discovery of transitional millisecond pulsars \citep[tMSPs, ][]{Archibald2009} which provide an insight into the end of this recycling process. tMSPs are MSPs that switch between a rotation-powered radio pulsar state and a sub-luminous LMXB state, and are progenitor systems to MSPs. There are currently only three confirmed tMSPs: PSR J1023+0038 \citep[hereafter J1023;][]{Archibald2009,Stappers2014}, IGR J18245-2452 also known as M28I in the globular cluster M28 \citep{Papitto2013}, and XSS J12270-4859 \citep{Bassa2014}. In addition to the three confirmed tMSPs, there are also a handful of candidate systems that display observational properties similar to tMSPs, but are yet to have an observed state change \citep[e.g.,][]{Bogdanov2015,Rea2017,Bahramian2018a,CotiZelati2019,Li2020,Miller2020}. It is worth noting that all three confirmed tMSPs are redback pulsar systems, MSPs with a companion of mass $0.2-0.5\ \textrm{M}_{\odot}$, which is being ablated by the pulsar wind \citep{Strader2019}.
    
    In the radio pulsar state tMSPs display typical redback pulsar properties. Radio pulsations can be detected \citep{Archibald2009, Papitto2013}, and the X-ray luminosity ($L_{\textrm{X}}$) from these sources is faint \citep[$L_{\textrm{X}}\lesssim10^{32}$ \si{\erg\per\second};][]{Linares2014}. In the sub-luminous disc state, these radio pulsations become undetectable as seen by \citet{Stappers2014}, and are instead replaced with bright, flat-spectrum radio continuum emission \citep{Hill2011,Deller2015}, although this emission has not been observed in this state for M28I. The X-ray emission in this state becomes brighter \citep[$L_{\textrm{X}}\sim10^{33}$ erg s$^{-1}$;][]{Bogdanov2015a}, and the gamma-ray flux of these systems has been observed to increase by at least a factor of five relative to the pulsar state \citep{Stappers2014,Johnson2015}. 
    
    A property of this low-level accretion state is the bimodal distribution of the X-ray flux of these systems, alternating between an X-ray high and low mode on a sub-hour time-scale, with occasional flares up to $L_{\textrm{X}}\sim10^{34}$ erg s$^{-1}$ in the high mode \citep{deMartino2013,Bogdanov2015a}. It was observed that there is a distinct anti-correlation between the radio and X-ray luminosity of J1023 during this state, which is thought to be due to the active pulsar \citep{Bogdanov2018}. Further studies of J1023 have also shown that it exhibits optical pulsations, which are possibly powered by the pulsar magnetosphere \citep{Ambrosino2017}.
    
    The anti-correlation between radio and X-ray luminosities also has consequences for placing tMSPs on the radio--X-ray luminosity plane. The radio--X-ray luminosity correlation between accreting BHs and NSs has been studied previously, and it is known that black hole X-ray binaries (BHXBs) display a higher radio luminosity compared to accreting NSs at similar X-ray luminosities \citep{Fender2001,Fender2003,Gallo2003,Maccarone2005,Migliari2006}. As such, by plotting a source's radio luminosity against its X-ray luminosity, some cautious guesses can be made as to the nature of the accreting compact companion. tMSPs on average are less radio bright than accreting BHs at the same X-ray luminosity \citep[e.g.,][]{Deller2015,Bogdanov2018,Papitto2013,deMartino2013,Bassa2014}, and lie below the accreting BH correlation of $L_{\textrm{R}} \propto L_{\textrm{X}}^{0.6}$ \citep{Gallo2014}. However, during a low X-ray mode, tMSPs can venture onto this track (as seen in the case of J1023 by \citealt{Bogdanov2018}), making it difficult to distinguish between an accreting BH and a tMSP based on radio and X-ray luminosity measurements alone. Additional multi-wavelength observations, such as radial velocity analyses of optical spectroscopic observations, help to alleviate this limitation.

\subsection{NGC~6652B}
    NGC 6652 is a globular cluster with a mass of \num{5.2e4} $\textrm{M}_{\odot}$ located at a distance of \SI{10}{kpc} \citep{Harris1996,Baumgardt2019}. It contains one bright LMXB (XB 1832-330 with $L_{\textrm{X}}\sim10^{36}$ erg s$^{-1}$), in addition to several other lower luminosity X-ray binaries \citep{Heinke2001, Stacey2012}. 
    
    NGC~6652B is the second brightest X-ray source in the cluster ($L_{\textrm{X}}\sim\SI{4e34}{\erg\per\second}$) and was first identified in optical studies of the cluster in 1998 by \citet{Deutsch1998} as star 49, a faint, blue object that was proposed as an optical counterpart to XB 1832-330. Subsequent \chandra observations of the cluster confirmed that it was not the optical counterpart to XB 1832-330, but rather the counterpart to a new, faint LMXB \citep[source B, ][]{Heinke2001}.
    
    Strong optical variability from the source has been observed \citep{Deutsch2000}, although the claimed periodicity of this variability ($\sim45$ minutes) has been disputed \citep{Heinke2001}. Observations by \citet{Engel2012} showed optical variations by up to 1 magnitude in the \textit{g'}-band on time-scales of hundreds of seconds. This strong blue variability also means that the apparent location of the optical counterpart on the main sequence is questionable.
    
    Significant X-ray variability over similar time-scales has also been observed, in addition to evidence of X-ray spectral variability \citep{Coomber2011,Stacey2012}. The measured values of the hydrogen column density ($N_{\textrm{H}}$) are consistent with the cluster foreground, and too low for obscuration by a disc to cause this spectral variation, with \citet{Stacey2012} suggesting that the cause of this variation is due to the propeller effect \citep{Illarionov1975}. This is a scenario that arises when the magnetic field of the neutron star creates a centrifugal barrier that prevents accretion onto the neutron star surface. NGC~6652B is loosely classified as a very faint X-ray transient (VFXT), defined as a transient with a peak X-ray luminosity $<10^{36}$ erg s$^{-1}$ \citep{Heinke2015}.
    
    In this paper, we present simultaneous \chandra and Karl G. Jansky Very Large Array (VLA) observations of this source taken as follow-up observations for the MAVERIC (Milky-way ATCA and VLA Exploration of Radio-sources in Clusters) survey \citep[Tudor et al., in prep.;][]{Shishkovsky2020,Bahramian2020}, in addition to simultaneous optical observations with Gemini North and optical spectroscopy obtained using the Southern Astrophysical Research Telescope (SOAR) 4 m telescope. Based on the hard X-ray emission, variable H $\alpha$ emission, and brighter than expected radio luminosity for a neutron star, we conclude that NGC~6652B is a tMSP candidate.

    This paper is organised as follows. \S~\ref{sec:Method} presents the data reduction and analysis methods, and the results are stated in \S~\ref{sec:Results}. \S~\ref{sec:Discussion} presents a discussion of the results and interpretations as to the nature of NGC~6652B.

\section{Observations and data reduction} \label{sec:Method}
    The MAVERIC survey consists of a deep radio continuum survey of 50 Galactic globular clusters performed using the ATCA and the VLA \citep[][Tudor et al., in prep.]{Tremou2018,Shishkovsky2020}. This is accompanied by a \textit{Chandra}/ACIS catalogue of X-ray sources in 38 Galactic globular clusters \citep{Bahramian2020}. 
    
    As part of this survey, the cluster NGC 6652 was observed by the ATCA for a total of 26.1 hr over three observing runs on 2015 April 19, 2015 June 21, and 2015 June 22 under the project code C2877. From these observations, a radio source at the position of the X-ray source NGC~6652B was identified (Tudor et al., in prep.). This source has an average 9 GHz radio flux density of $76\pm4$\ \si{\micro Jy}, and an inverted spectral index of $\alpha=0.38\pm0.13$ ($S_{\nu}\propto\nu^{\alpha}$). To further investigate this source, we performed simultaneous radio and X-ray observations using the VLA and the \chandra X-ray Observatory. The data used in this paper are listed in Table~\ref{tab:data}, and in the following sections we describe the data reduction.
    
    \begin{table}
    \centering
    \begin{tabular}{llcc}
        \hline
        \hline
        Observatory/Inst. & Obs. ID & Date & Duration\\
        \hline
        VLA (B configuration) & 16A-325 & 2016 May 27 & 0.75 hr\\
                              &         & 2016 May 29 & 0.75 hr\\
                              &         & 2016 June 08 & 0.75 hr\\
                              &         & 2016 June 11 & 0.67 hr\\
        VLA (C configuration) & SI0399 & 2017 May 22 & 2.75 hr\\
        \chandra/ACIS & 18987 & 2017 May 22 & 10 ks \\
        Gemini-North/AcqCam & GN-2017A-DD-7 & 2017 May 22 & 0.92 hr\\
        SOAR/Goodman & - & 2015 May 24 & 0.5 hr\\
         & & 2015 May 28 & 1.83 hr\\
         & & 2015 Aug 08 & 0.67 hr\\
        \hline
    \end{tabular}
    \caption{The multi-wavelength data used in this study.}
    \label{tab:data}
    \end{table}

\subsection{Follow-up radio data} \label{sec:radio}
    We observed NGC~6652B a total of five times with the VLA over the course of a year under two different project codes. The first set of observations under project code 16A-325 consists of four epochs of observations: three 0.75 hour observations on 2016 May 27, 2016 May 29, and 2016 June 8, and one 0.67 hour observation on 2016 June 11. The array was in the B configuration for these observations, and the target source was observed at C band with two 2 GHz basebands centred at 5.0 GHz and 7.0 GHz respectively. The flux density and bandpass calibrator for these observations was 3C48, and the complex gain calibrator was J1820-2528.
    
    In addition to these four epochs, we observed NGC~6652B with the VLA on 2017 May 22 from 08:09-11:32 UTC under the project code SI0399, giving an on-target time of 2.75 hours. This observation was performed simultaneously with X-ray observations of the source made using the \chandra X-ray Observatory. The array was in the C configuration, and the target source was observed at X band with two 2 GHz basebands centred at 9.0 and 11.0 GHz, respectively. 3C286 was used as the flux density and bandpass calibrator, and J1820-2528 was used as the complex gain calibrator. 
    
    The data were imported into the Common Astronomy Software Application \citep[\textsc{casa}; ][]{McMullin2007} for data reduction and imaging. The data were reduced using the VLA \textsc{casa} calibration pipeline version 5.6.2-3, which automatically flags and calibrates the data. Any remaining spurious data were manually flagged from the dataset before the calibration pipeline was re-run. Imaging was performed with the \verb#tclean# task using Briggs weighting with a robust parameter of 0 to balance sensitivity and resolution. The flux density and position of the target source were determined using the \textsc{casa} task \verb#imfit#. The rms, clean beam size, and measured flux densities of NGC~6652B are shown in Table~\ref{tab:radio_obs}. This image is shown in Figure~\ref{fig:images}.

    \begin{table}
    \centering
    \begin{tabular}{llccccc}
        \hline
        \hline
        Date & Band & $B_{\textrm{maj}}$ & $B_{\textrm{min}}$ & $B_{\textrm{pa}}$ & rms & Flux density\\
             &      & (\si{\arcsecond}) & (\si{\arcsecond}) & (\si{\degree}) & (\si{\micro Jy \per bm}) & (\si{\micro Jy})\\
        \hline
        2016 May 27 & C & 2.81 & 0.73 & 21.98 & 9.41 & $46.3\pm6.9$\\
        2016 May 29 & C & 2.89 & 0.74 & 23.72 & 8.04 & $39.6\pm5.8$\\
        2016 June 08 & C & 2.66 & 0.73 & 19.49 & 8.81 & $69.9\pm6.6$\\
        2016 June 11 & C & 2.50 & 0.75 & 17.62 & 8.82 & $88.4\pm6.8$\\
        2017 May 22 & X & 3.89 & 1.38 & 1.51 & 3.73 & $79.2\pm2.2$\\
        \hline
    \end{tabular}
    \caption{The FWHM of the clean beam, consisting of the semi-major axis ($B_{\textrm{maj}}$), the semi-minor axis ($B_{\textrm{min}}$), and position angle ($B_{\textrm{pa}}$), and the central rms of the five VLA observations observations. The flux density of NGC~6652B and the frequency band of each observation are also listed. C band corresponds to 4-8 GHz and X band corresponds to 8-12 GHz.}
    \label{tab:radio_obs}
    \end{table}    

\subsection{X-ray data}
    NGC~6652B was observed with the \textit{Chandra} X-ray Observatory on 2017 May 22 from 08:14-11:43 UTC during observing cycle 18 under the observation ID 18987, giving an exposure time of 10 ks. This observation was simultaneous with a radio observation with the VLA (\S~\ref{sec:radio}). The observations were performed in Faint mode using ACIS-S and the target was positioned on chip S3 using a custom subarray of 128 rows to reduce pileup.
    
    For data processing and analysis, we used \textsc{ciao} 4.12.1 and \textsc{caldb} 4.9.3 \citep{Fruscione2006}. The 0.3-10 keV image is shown in Figure~\ref{fig:images}. The \chandra data were reprocessed using \verb#chandra_repro#, and \verb#dmextract# was used to extract a light curve binned at \SI{200}{\second} from the data in the 0.3-10 keV energy range. Source and background spectra were extracted using \verb#specextract#, with source and background regions chosen to avoid contamination by the bright LMXB XB 1832-330. A circular extraction region with a radius of \SI{2}{\arcsecond} was used for the source, and an annulus with an inner radius of \SI{2.5}{\arcsecond} and an outer radius of \SI{7}{\arcsecond} was used for the background. Spectral analysis was performed using \textsc{xspec} 12.11 \citep{Arnaud1996}.
    
\subsection{Optical data}
\subsubsection{Photometry}
\begin{figure*}
        \centering
        \includegraphics[scale=0.5, width=0.48\linewidth]{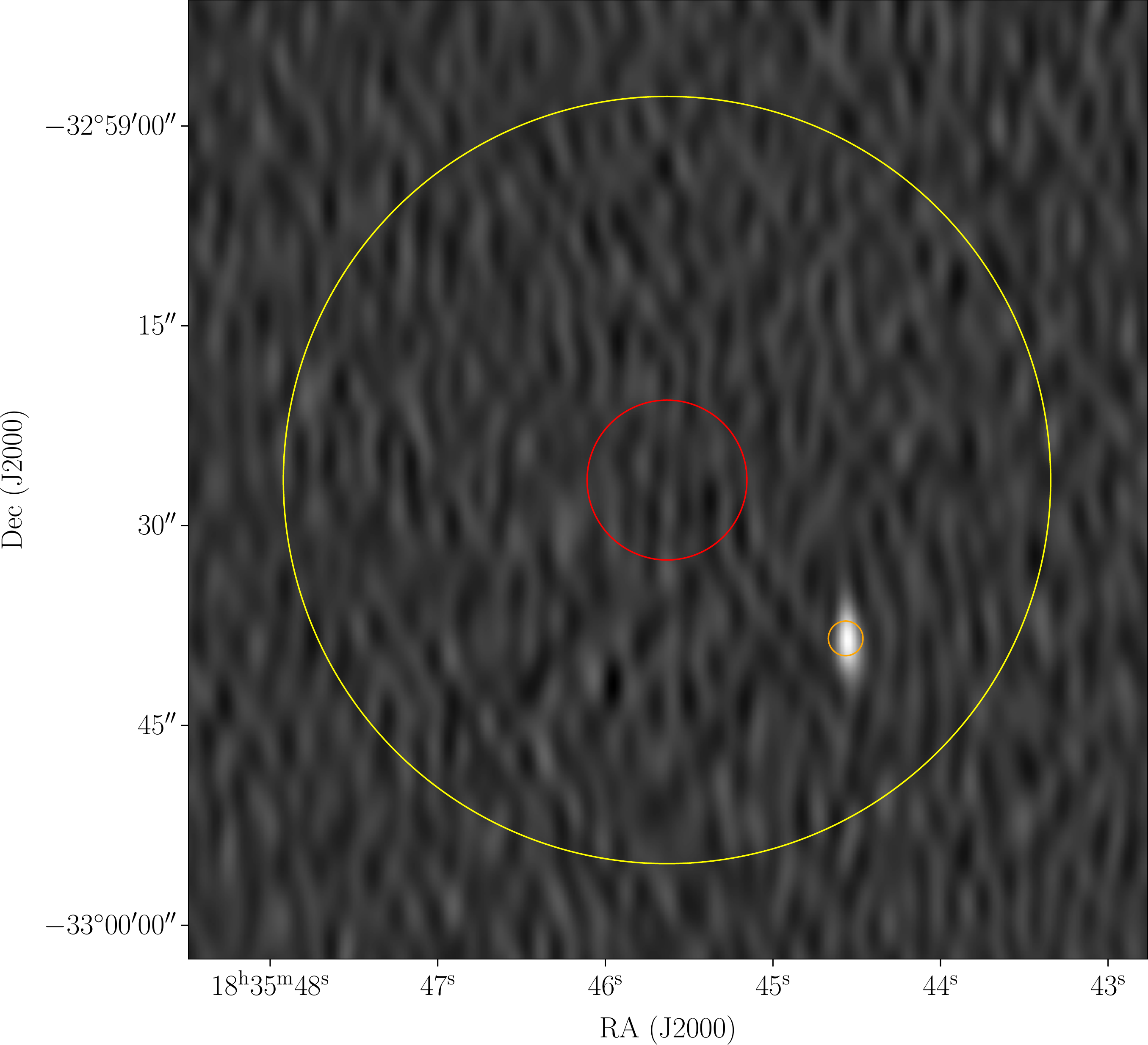}
        \includegraphics[scale=0.5, width=0.48\linewidth]{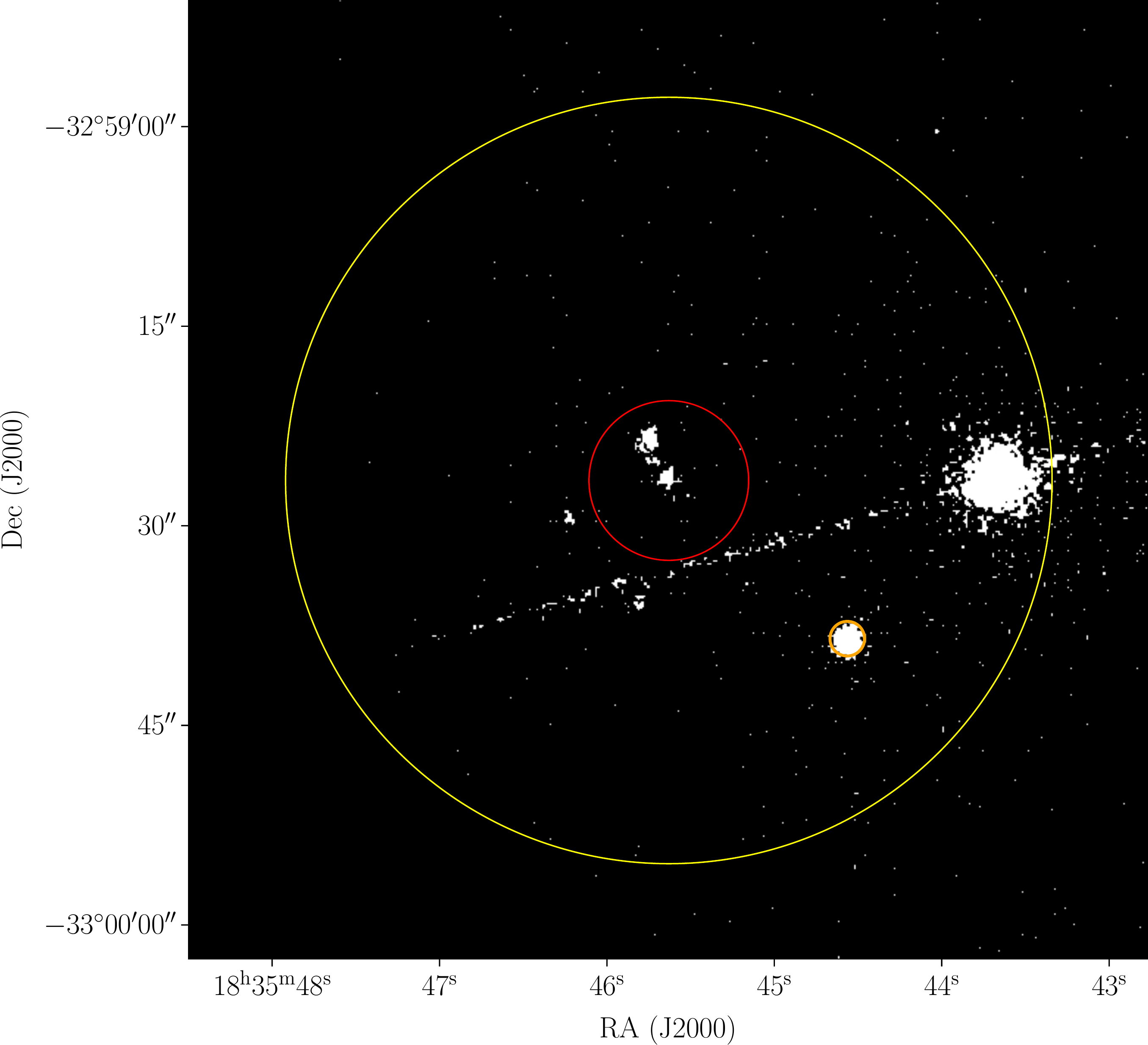}
        \includegraphics[scale=0.5, width=0.48\linewidth]{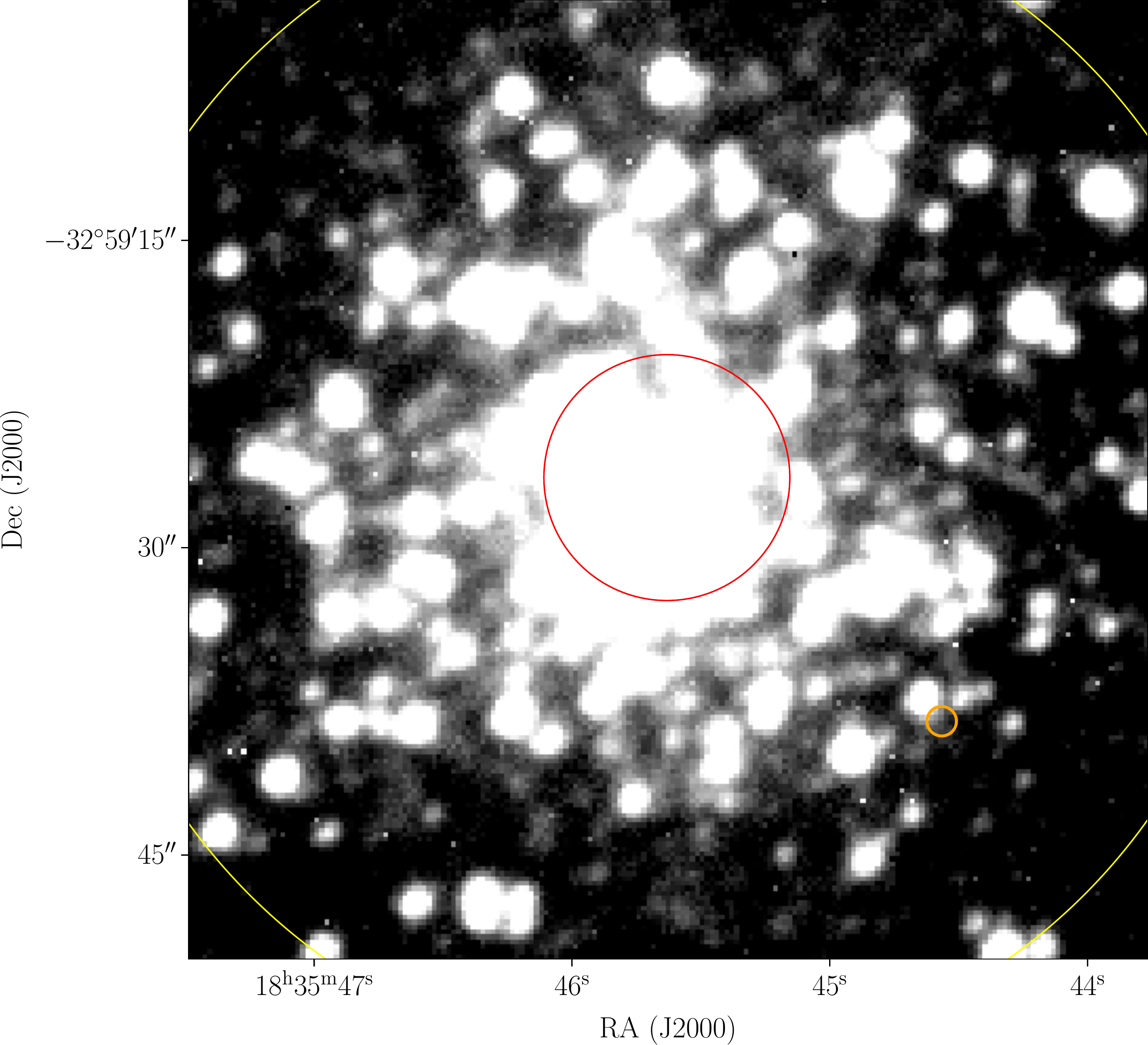}
        \caption{The 10 GHz VLA image (top left), the 0.3-10 keV \chandra image (top right), and the \textit{V} filter Gemini image (bottom) of NGC 6652B and the surrounding field. NGC~6652B is highlighted by the orange circle. The red and yellow circles indicate the core and half-light radii of the cluster NGC 6652 respectively.}
        \label{fig:images}
    \end{figure*}

    Optical observations of NGC~6652B were obtained with the Gemini North Acquisition Camera on 2017 May 22 from 10:35-11:30 UTC, giving 55 minutes of observations that were simultaneous with the VLA and \textit{Chandra}. The observations were taken with the \textit{V} filter under the project code GN-2017A-DD-7, and consisted of 1332 \SI{2}{s} snapshots for the purpose of fast photometry. The data were reduced using the \textsc{python} package \textsc{ccdproc} \citep{ccdproc} and astrometrically aligned for stacking and further analysis. For fast photometry and searching for variations, we used the software package \textsc{isis} 2.2 \citep{Alard1998,Alard2000}. This image is shown in Figure~\ref{fig:images}.
    
\subsubsection{Spectroscopy}
    We obtained spectra of NGC~6652B using SOAR/Goodman on three different nights: 2015 May 24, 2015 May 28, and 2015 Aug 18. All observations used a 400 l mm$^{-1}$ grating and a 0.84\arcsec\ longslit, yielding a resolution of about 5.0 \AA\ full-width at half-maximum (FWHM) over a nominal wavelength range $\sim 3000$--7000 \AA\ (though the signal below $\sim 3800$ \AA\ is very low). On 2015 May 24 we obtained a pair of 15-min exposures. On 2015 May 28 we first obtained a pair of 15-min exposures before taking two pairs of 20-min exposures, for six spectra total during this night. We obtained a final pair of 20-min exposures on 2015 Aug 18.

    The spectra were all reduced and optimally extracted in the usual manner, and wavelength-calibrated using FeAr arc lamp exposures obtained immediately following the science spectra. Heliocentric velocity corrections were calculated using \textsc{iraf} \citep{Tody1986,Tody1993} using the package \textsc{rvcorrect} and applied to the spectra. We searched for variability between the spectra and also stacked them to increase the signal-to-noise ratio for possible features originating from the accretion disc.
    
    \begin{figure*}
        \centering
        \includegraphics[scale=0.4]{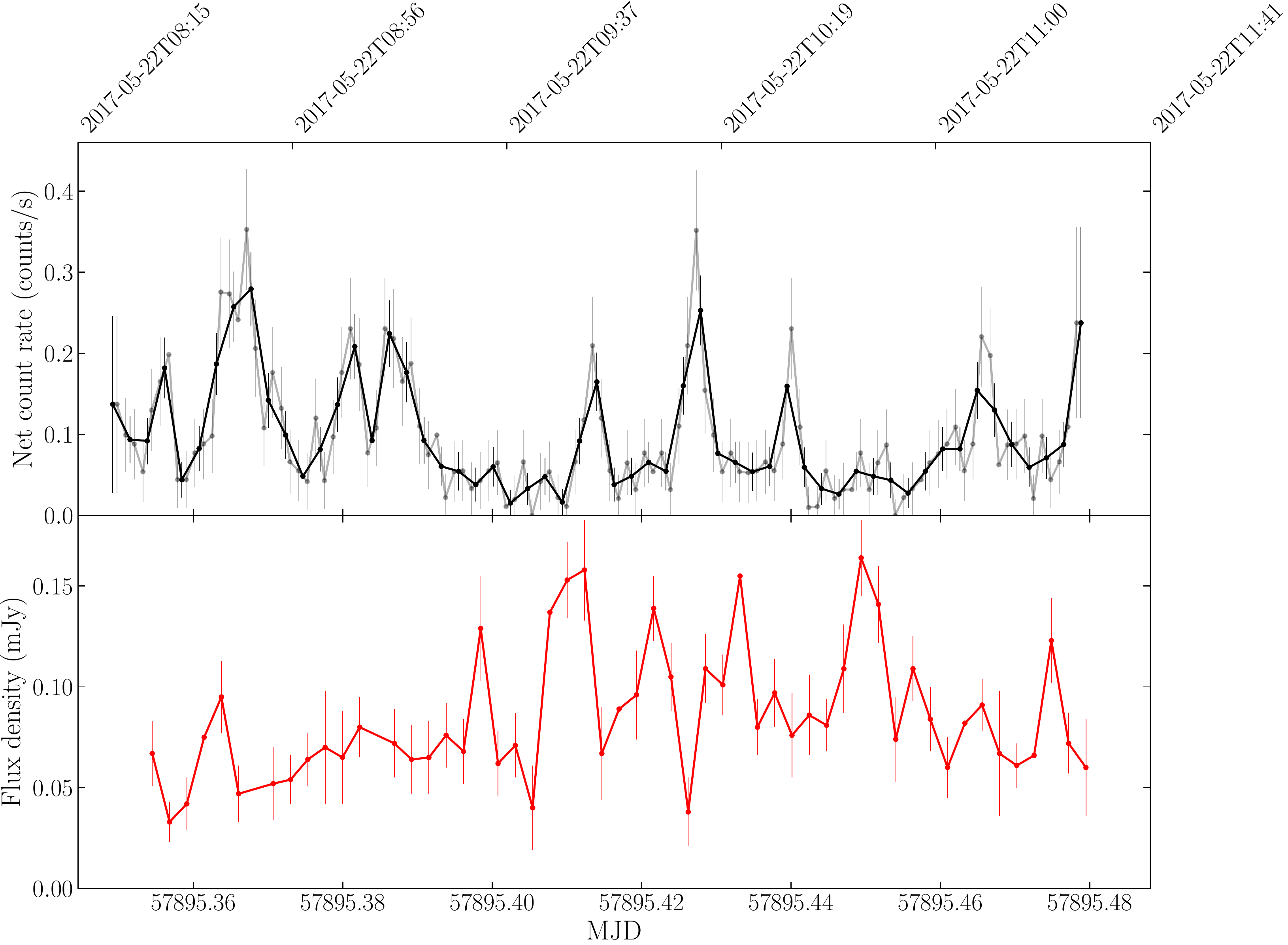}
        \caption{\textit{Top}: The \chandra background subtracted light curve extracted in the 0.3-10 keV energy band with a binning of \SI{200}{\second}. This light curve is shown in black. The grey light curve shows the same data with a binning of \SI{100}{\second}. Numerous X-ray flares are seen over the duration of the observation. \textit{Bottom}: The 8-12 GHz VLA light curve with a binning of \SI{200}{\second}. Some radio flares are visible in the later half of the observation.}
        \label{fig:chandra_vla}
    \end{figure*}

\section{Analysis and results} \label{sec:Results}
\subsection{Searching for correlated variability in the radio and X-ray bands}
    To investigate whether there is any correlated variability between the radio and X-ray bands in NGC~6652B, we extracted a background subtracted light curve with a binning of \SI{200}{\second} in the 0.3-10 keV energy range from the \chandra X-ray data. This is shown in the top panel of Figure~\ref{fig:chandra_vla}. The source shows variability over a time-scale of a few hours, with several flaring events with count rates $>0.1$ \si{counts\per\second} present in the light curve. The fractional RMS variability amplitude 
    for the X-ray light curve is 0.598, implying that the X-ray emission from the source is variable. The average 1-10 keV X-ray flux during this observation is $1.5\pm0.1 \times10^{-12}$ \si{\erg\per\second\per\square\centi\metre}.
    
    The bottom panel of Figure~\ref{fig:chandra_vla} shows the light curve extracted from the simultaneous VLA radio observation. This observation was split into several bins with a width of \SI{200}{\second} to match the timing resolution of the \chandra light curve and each bin was imaged. The task \verb#imfit# was used to extract the flux density from these images. The fractional RMS variability for the radio light curve is 0.316, implying moderate variability. The average 10 GHz radio flux density of the source during this observation is $79.2\pm2.2$ \si{\micro Jy}.

    \begin{figure}
        \centering
        \includegraphics[width=\columnwidth]{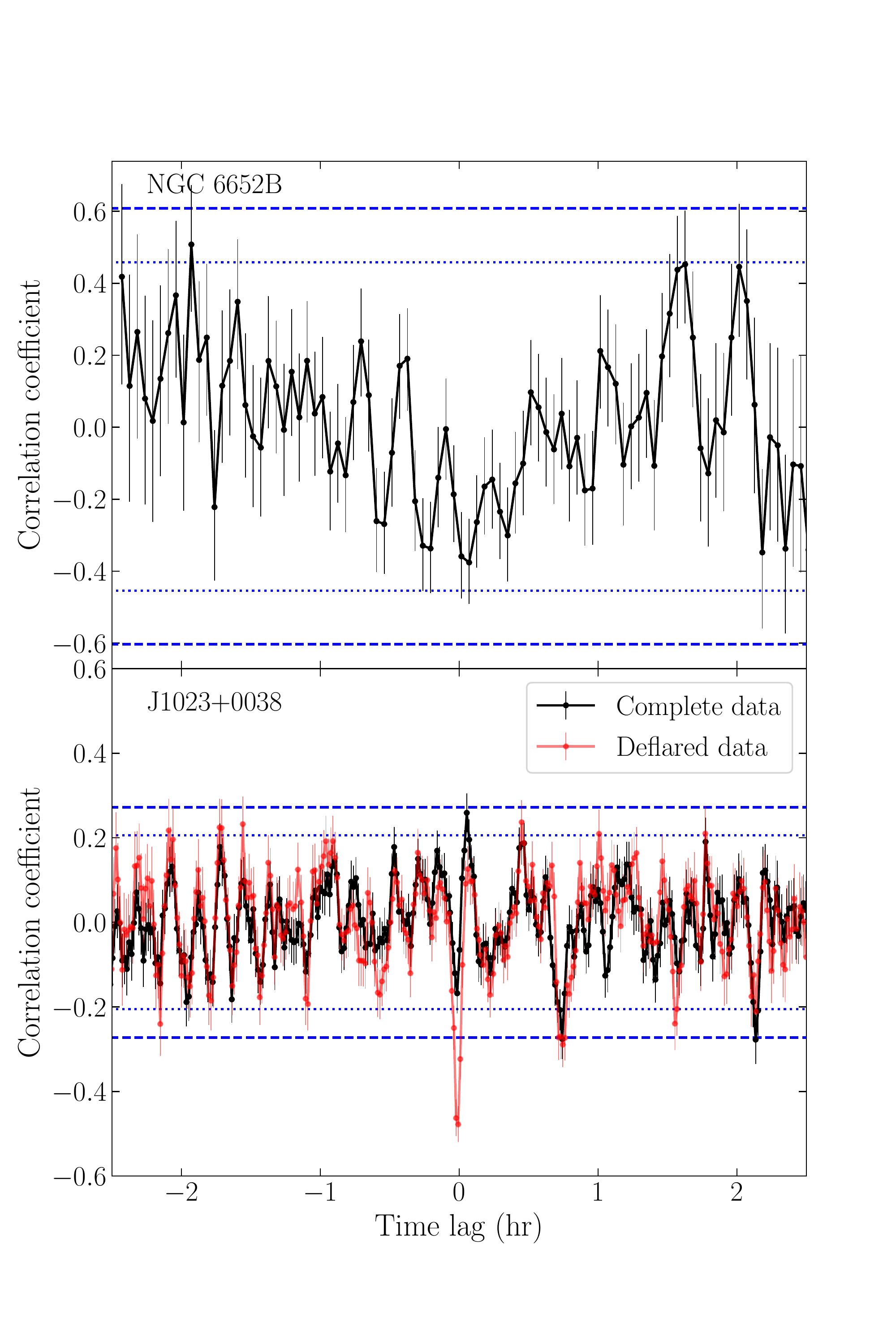}
        \caption{\textit{Top}: The Z-transformed discrete correlation function of the radio and X-ray light curves of NGC~6652B. The 95\% and 99\% confidence intervals are shown as blue dotted and blue dashed lines respectively. No clear correlation is seen to $>99\%$ confidence. \textit{Bottom}: The Z-transformed discrete correlation function of the radio and X-ray light curves of J1023+0038 for two cases. The black line indicates the correlation function for the complete dataset, where the significant radio and X-ray flares seen in the light curves have been included in the dataset. In this complete dataset, no significant anti-correlation is seen between the radio and X-ray light curves. The red line indicates the correlation function for the dataset where the radio and X-ray flares have been removed, to leave only the high and low mode data. The blue dotted and dashed lines indicate the 95\% and 99\% confidence intervals respectively for this ``deflared'' dataset. With the flares removed from the dataset, anti-correlation is seen to $>99\%$ confidence between the radio and X-ray light curves in the high and low mode data. This is discussed further in \S~\ref{sec:tMSP_candidate}.}
        \label{fig:cross-correlation}
    \end{figure}
    
    We calculated the Z-transformed discrete correlation function\footnote{\url{https://webhome.weizmann.ac.il/home/tal/zdcf2.html}} \citep[ZDCF;][]{Alexander1997} of the two light curves to determine if there was any (anti-)correlation between them. The ZDCF was performed with a 200 s time bin. This cross-correlation function is shown in the top panel of Figure~\ref{fig:cross-correlation}. To estimate the uncertainty on the cross-correlation function, we simulated 1000 light curves with the same power spectrum as the data, and calculated the ZDCF for each pair of simulated light curves. From this, we determined the 95\% and 99\% confidence intervals on the cross-correlation function. This showed that there is no clear correlation between the simultaneous radio and X-ray light curves to $>99\%$ confidence. 

\begin{figure*}
        \centering
        \includegraphics[scale=0.5]{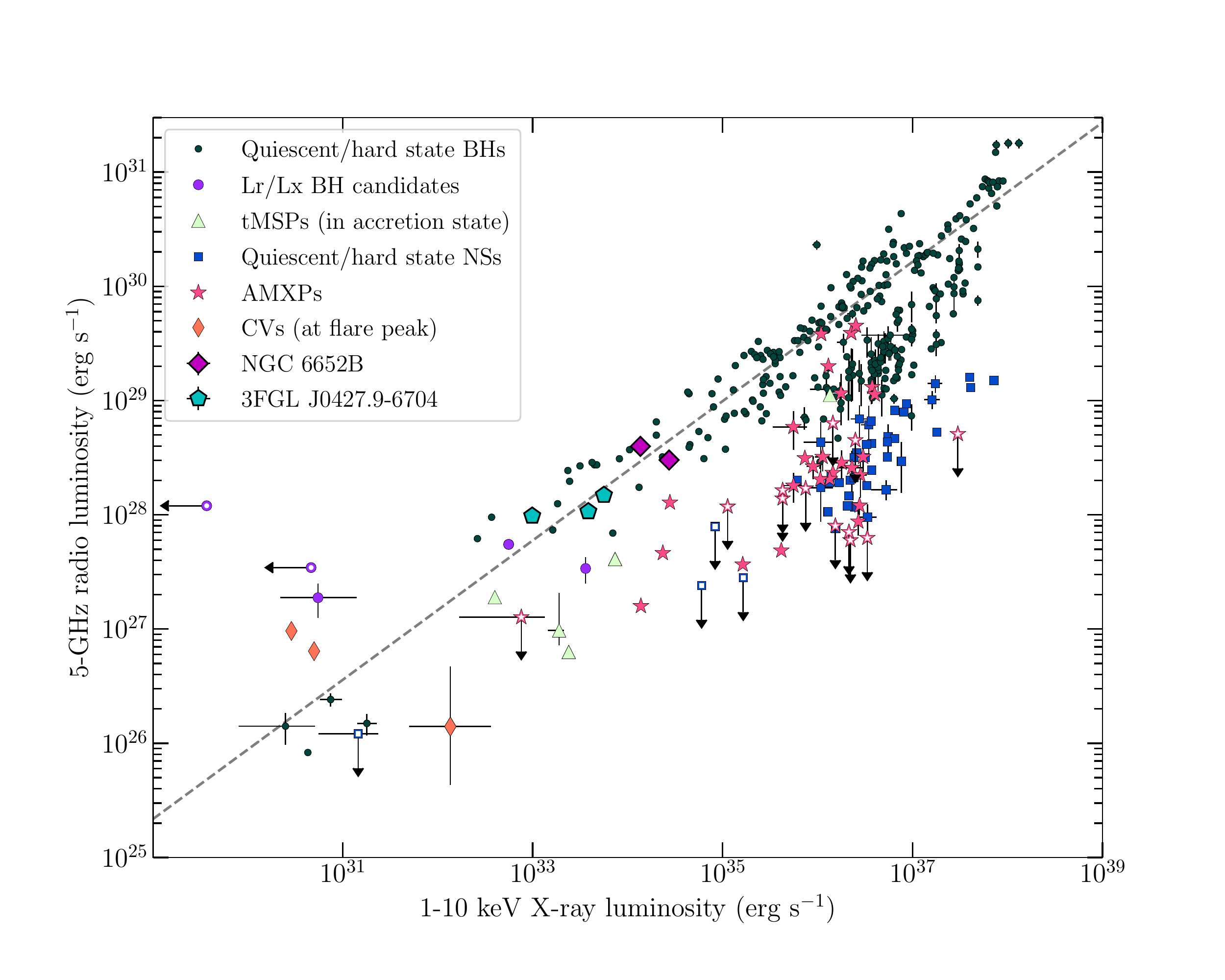}
        \caption{The radio--X-ray correlation for accreting systems, with the high and low X-ray count rate regimes of NGC~6652B plotted. Note the error bars are too small to be seen. The different coloured markers represent different types of accreting binaries from the database of \citet{Bahramian2018}, and the dashed line represents the correlation for accreting BHs \citep{Gallo2014}. NGC~6652B sits near the BH correlation, with this position changing for different X-ray count rate regimes. This behaviour is different to that of J0427.}
        \label{fig:lrlx}
    \end{figure*}

    To investigate the behaviour of the source both during and outside an X-ray flaring event, we binned the X-ray and radio data based on periods of high (>\SI{0.09}{counts\per\second}) and low ($\leq0.09$ \si{counts\per\second}) X-ray counts. From this we calculated the corresponding X-ray fluxes, using X-ray spectroscopy (see \S~\ref{sec:xray_spec}), and radio flux densities for these time periods. These results are shown in Table~\ref{tab:x-ray_spec}. 
    By assuming the distance to the cluster NGC 6652 to be \SI{10}{kpc} \citep{Harris1996} and using a radio spectral index of $0.38\pm0.13$ for NGC~6652B, as measured in the original MAVERIC survey (Tudor et al., in prep), the 5 GHz radio luminosities and 1-10 \si{keV} X-ray luminosities were calculated, and the source was plotted on the radio--X-ray plane for accreting systems for different X-ray count rates (Figure~\ref{fig:lrlx}). These radio and X-ray luminosities place NGC~6652B near the radio--X-ray correlation for accreting BHs \citep{Gallo2014}. Furthermore, NGC~6652B appears to show enhanced radio luminosity during periods of low X-ray counts and vice versa during our observations, however this behaviour only appears when considering the average flux densities during periods of high and low X-ray counts, that is the values shown in Table~\ref{tab:x-ray_spec}. However, as shown through the cross-correlation analysis, there is no correlation to $>99\%$ confidence in the 200 s binned radio and X-ray light curves.

\subsection{Longer term radio behaviour}
    The additional 4-8 GHz VLA observations taken a year before the simultaneous observations provide an opportunity to study the longer term radio behaviour of NGC~6652B. The measured 9 GHz flux density of the source in the original MAVERIC survey data from 2015 is $76\pm4$ \si{\micro Jy}. The flux densities from our 4-8 GHz observations and 8-12 GHz observations were measured using the \textsc{CASA} task \textsc{IMFIT} and are shown in Table~\ref{tab:radio_obs}. The four 4-8 GHz observations show that the 6 GHz flux density of the source ranges between 40 and \SI{88}{\micro Jy} over the course of two weeks. The 8-12 GHz observation taken a year later showed the source with a 10 GHz flux density of $79.2\pm2.2$ \si{\micro Jy}. Only moderate variations by a factor of $\sim2$ are seen, at similar levels to the variation seen within the radio light curve of the X band observation. 

\subsection{X-ray spectroscopy} \label{sec:xray_spec}
    X-ray spectra of NGC~6652B were extracted for intervals of high (>\SI{0.09}{counts\per\second}) and low ($\leq0.09$ \si{counts\per\second}) X-ray counts, and for the entire observation using \verb#specextract#, with the source spectra binned to have 20 counts per bin. \textsc{xspec} 12.11 \citep{Arnaud1996} was used for spectral analysis, assuming \citet{Wilms2000} abundance of elements and $\chi^2$ statistics for inference. Previous X-ray spectral analysis of NGC 6652B has been done in-depth by \citet{Stacey2012}. From a 47.5 ks \chandra observation, a variety of spectral models were fit to the spectrum of NGC 6652B, and it was found that a power-law model is a good fit to the data. In light of this, we preliminarily fit each spectrum with an absorbed power law model (\verb#tbabs# $\times$ \verb#pegpwrlw#), which indicated varying values of the hydrogen column density ($N_{\textrm{H}}$) between high and low X-ray counts, suggesting the presence of intrinsic absorption in the system. We also investigated whether our data were affected by pileup. For a $1/8$ subarray we calculated the pileup to be $<5\%$, and verified this by including pileup in our spectral models, where it had a minimal impact on the fitted parameters.
    
    \begin{table*}
    \begin{tabular}{cccccccc}
        \hline
        \hline
        X-ray count rate threshold & Radio flux density & Galactic $N_{\textrm{H}}$ & Intrinsic $N_{\textrm{H}}$ & $\Gamma$ & Unabsorbed flux & $\chi^2_{\nu}$/d.o.f. & N.H.P. \\
        (counts/s)                 & (\si{\micro Jy}) & (\SI{e22}{\per\square\centi\metre}) & (\SI{e22}{\per\square\centi\metre}) & & (\SI{e-12}{\erg\per\second\per\square\centi\metre}) &  & \\
        \hline
        $>0.09$      & $65.6\pm3.8$ & (0.078) & $< 0.086^*$ & $1.45^{+0.15}_{-0.13}$ & $2.29^{+0.19}_{-0.24}$ & 0.70/26 & 0.868             \\
        $\leq0.09$   & $86.4\pm2.7$ & (0.078) & $< 0.611^*$ & $1.13^{+0.37}_{-0.33}$ & $1.14^{+0.22}_{-0.19}$ & 0.38/13 & 0.976             \\
        Average      & $79.2\pm2.2$ & (0.078) & $< 0.166^*$ & $1.32^{+0.17}_{-0.12}$ & $1.52^{+0.14}_{-0.14}$ & 0.72/41 & 0.912             \\
        \hline
    \end{tabular}
    \caption{The results of the X-ray spectral fitting and the measured radio flux densities in the different X-ray count rate regimes. The radio flux densities are measured at 10 GHz, and the radio uncertainties reported are the $1\sigma$ values. All other uncertainties are reported at 90\% confidence for a single parameter of interest. The values in parentheses indicated values that were frozen when fitting. The absorption column indicates the second absorption parameter included to test for intrinsic absorption, and the * indicates a 2.706 $\Delta\chi^2$ upper limit to these values (equivalent to a 90\% confidence region for a single parameter of interest). The unabsorbed flux is in the 1-10 keV band. N.H.P. is the null hypothesis probability of the fit, where the null hypothesis is that the data is drawn from the model, meaning a higher probability is better.}
    \label{tab:x-ray_spec}
    \end{table*}
    
    To investigate the change in column density, we fit the high and low X-ray count rate spectra with an absorbed power law model with an additional absorption parameter to account for a possible intrinsic absorption (\verb#tbabs# $\times$ \verb#tbabs# $\times$ \verb#pegpwrlw#). Foreground reddening in the direction of NGC 6652 is $E(B-V)=0.09$ \citep{Harris1996}, and using the relations calculated by \citet{Bahramian2015} and \citet{Foight2016}, we estimated the column density to the cluster to be \SI{7.8e20}{\per\square\centi\metre}. We set the first absorption component in the model to this value (representing the cluster foreground absorption), and allowed the second absorption parameter to vary to test the presence of intrinsic absorption. These results are shown in Table~\ref{tab:x-ray_spec}.  
    
\begin{figure}
        \centering
        \includegraphics[width=\columnwidth]{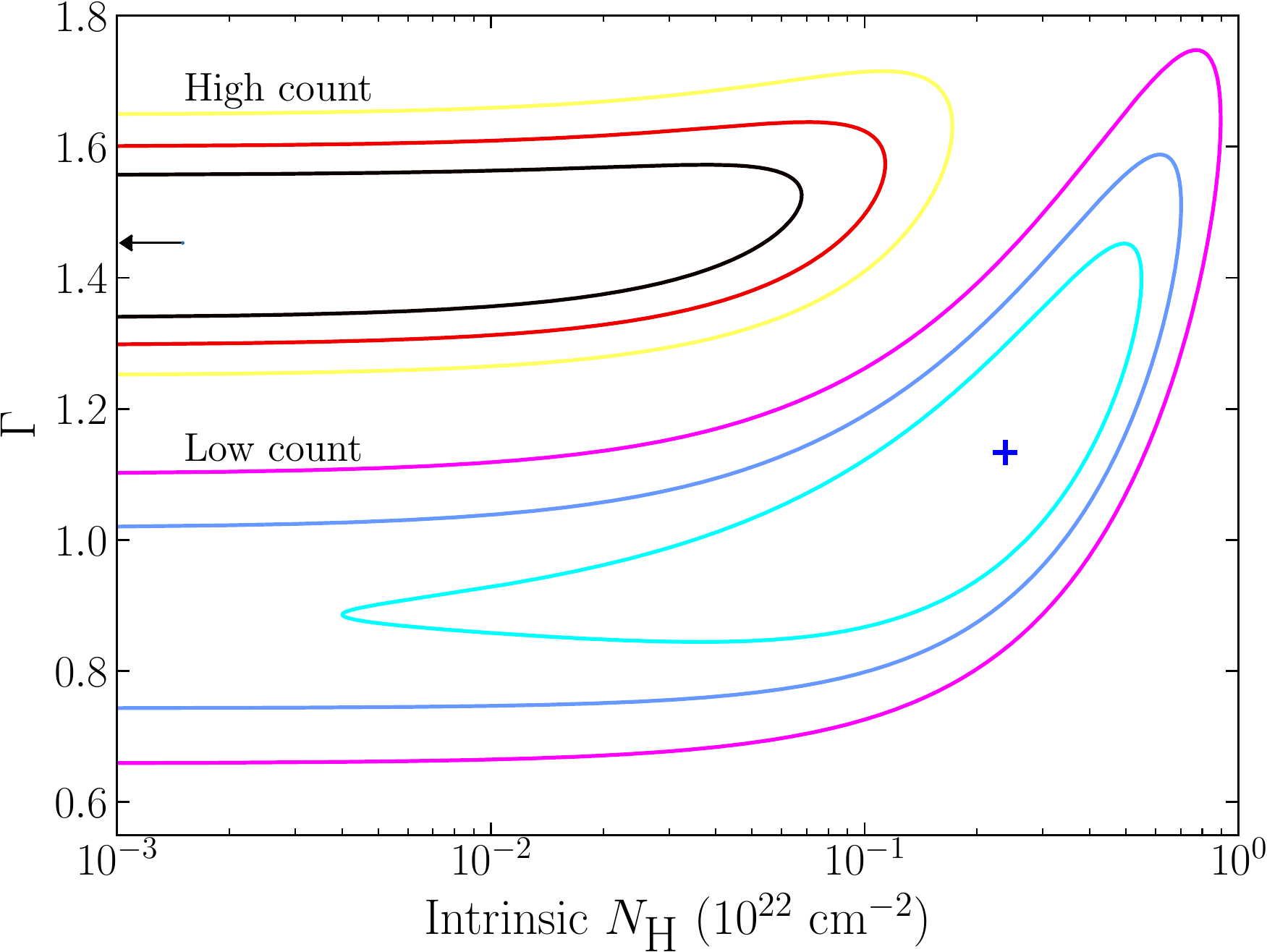}
        \caption{The contour plots for the high count (top left) and low count (bottom right) rate X-ray spectral fits. The different colour maps indicate that the contours are for two different datasets. The plot indicates how the $\chi^2$ of the fit changes as the parameter space of the photon index and intrinsic column density is explored. The black arrow indicates that the best-fit parameters for the high X-ray count spectrum is beyond the x-axis limit of the plot, and the blue cross indicates the best-fit parameters for the low X-ray count rate spectrum. The contours indicate $1\sigma$, $2\sigma$, and $3\sigma$ levels for a single parameter of interest. The two fits are inconsistent with each other, and the intrinsic column density for the low X-ray count rate spectrum is inconsistent with zero at a $\geq1\sigma$ level. This indicates there is some evidence for intrinsic absorption in the system during these periods of low X-ray counts.}
        \label{fig:x-ray_contour}
    \end{figure}

    The spectra are hard ($\Gamma\sim1.3$), indicating non-thermal emission is present, with the low count rate spectrum indicating slightly harder emission ($\Gamma\sim1.1$). The measured photon indices are consistent within uncertainties. While the values of our second absorption parameter are consistent with zero for both the high and low X-ray count rates, we note that to a $\geq1~\sigma$ level the low count rate value is inconsistent with zero, indicating that there is marginal evidence for intrinsic absorption. This is highlighted in the contour plots for the high and low count rate spectra, shown in Figure~\ref{fig:x-ray_contour}. 
    
\subsection{Optical spectroscopy}

    \begin{figure*}
        \centering
        \includegraphics[width=0.95\textwidth]{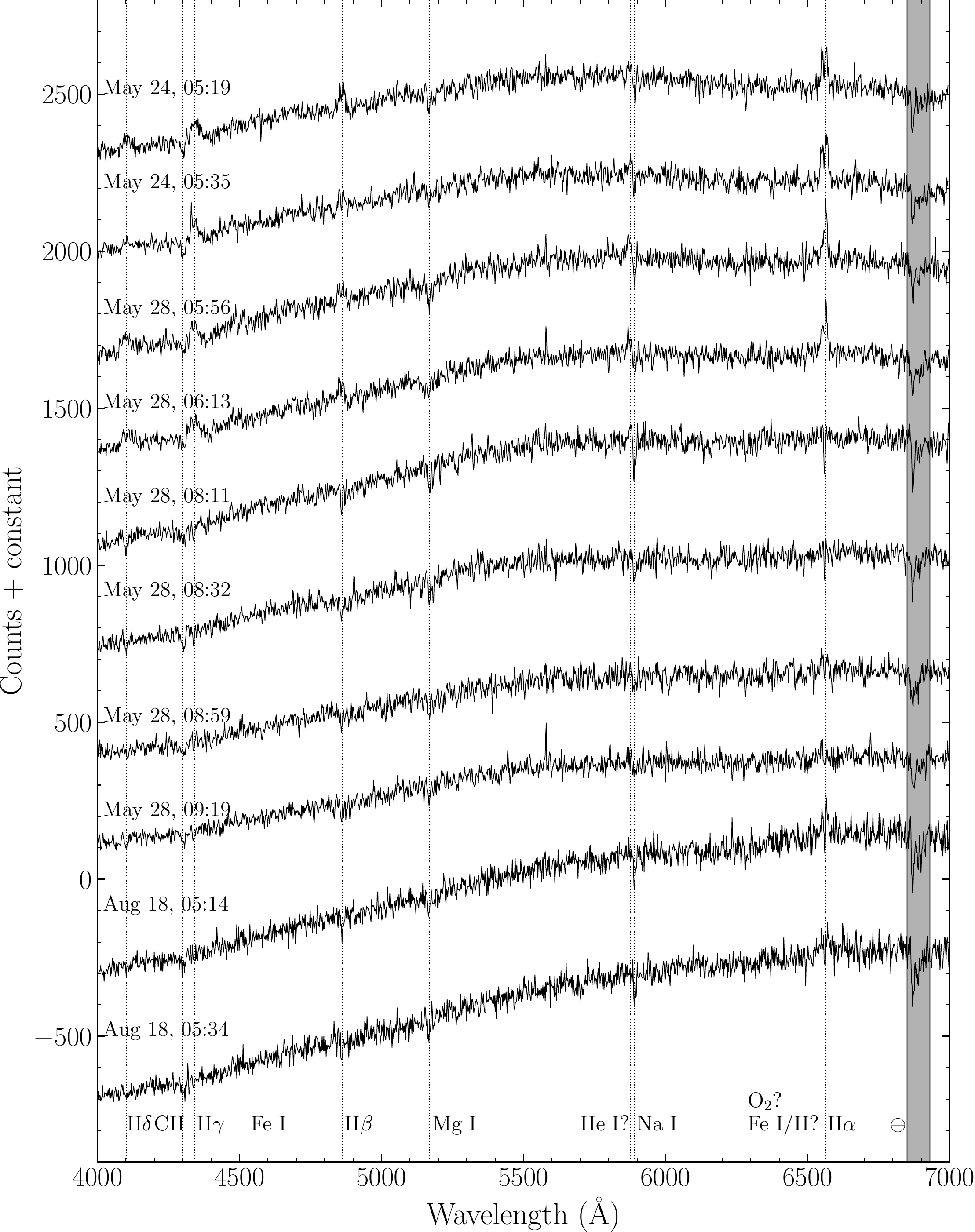}
        \caption{The ten optical spectra obtained of NGC~6652B. Several spectral features are shown with dashed lines. The first four Balmer features are seen, with prominent double-peaked H $\alpha$ emission visible in addition to H $\beta$, H $\gamma$, and H $\delta$ emission features seen in the first four observations before switching to absorption. Several stellar absorption features are visible, such as the G-band due to the CH molecule seen at $\sim\SI{4300}{\angstrom}$, and various other metallic absorption features such as Fe and Mg absorption lines. The shaded region indicates telluric absorption features.}
        \label{fig:soar_stacked}
    \end{figure*}
    
    We obtained ten optical spectra of NGC~6652B using the Goodman spectrograph on the SOAR 4 m telescope. These spectra are shown in Figure~\ref{fig:soar_stacked}. An inspection of these spectra shows some stellar absorption features and some variable emission features typical of accreting X-ray binary systems. At $\sim\SI{4300}{\angstrom}$ the G-band absorption feature is seen. The G-band is an absorption band due to the CH molecule, and is a feature of early F to early K-type stars. Other metallic absorption features are seen at $\sim\SI{4530}{\angstrom}$ (Fe \textsc{i}), $\sim\SI{5180}{\angstrom}$ (Mg \textsc{i}) and $\sim\SI{5900}{\angstrom}$ (Na \textsc{i} doublet). The absorption line at $\sim\SI{6290}{\angstrom}$ is potentially due to molecular oxygen from the Earth's atmosphere, with further telluric absorption features seen from \SI{6850}{\angstrom} to \SI{6930}{\angstrom}. The first four Balmer features are seen and there is evidence of broad He \textsc{i} emission seen at $\sim\SI{5875}{\angstrom}$ in some of the spectra. The presence of the G-band, and the indicated metallic absorption lines are consistent with an early to mid G-type star (see \S~\ref{sec:optical_counterpart}), however robust identification of the optical counterpart is difficult due to potential contamination from the crowded region around the source.
    
    Broad H $\alpha$ emission lines are prominent in the first four spectra, indicating that this source is most likely an accreting binary system \citep{Casares2015,Casares2016}. While for bona fide quiescent X-ray binaries, the relations of \citet{Casares2015,Casares2016} can be used to estimate the projected radial velocity amplitude and the binary mass ratios using emission lines, these correlations have not been established to work for brighter objects, and do rely on the ionisation profile of the accretion disc being relatively similar to that of a quiescent system's accretion disc.  Furthermore, the mass ratio estimates require averaging carefully over orbital phase, and with relatively strong variability, as we see for NGC~6652B, far more spectra are needed to give an appropriately weighted average of the data. Thus, this is not an appropriate situation for these correlations to be implemented.
    
\begin{figure}
        \centering
        \includegraphics[width=\columnwidth]{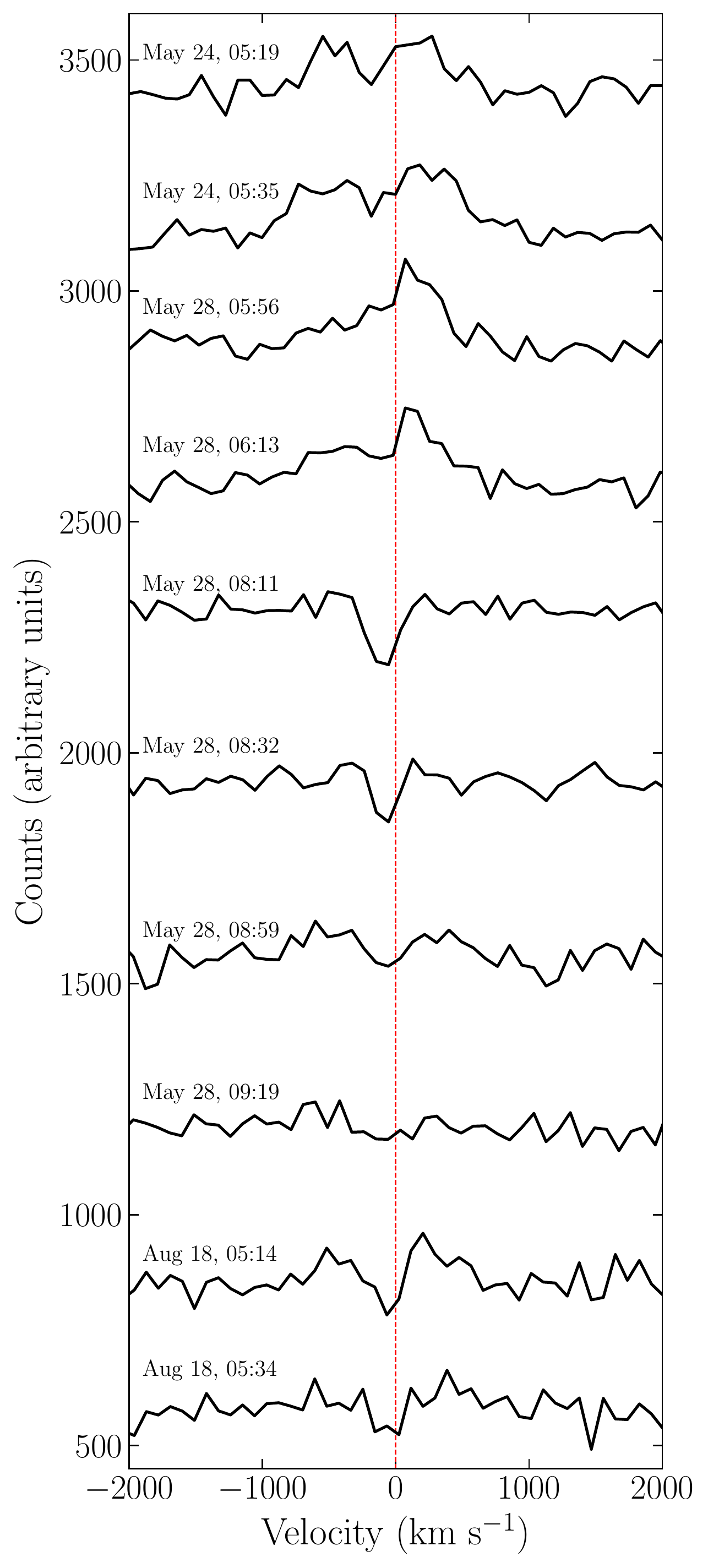}
        \caption{The variation in the H $\alpha$ line over ten spectra of NGC~6652B. The red dashed line indicates a velocity of \SI{0}{\kilo\metre\per\second}. The H $\alpha$ feature goes from double-peaked emission in the first observation to a single emission feature, before transitioning to full absorption in the fifth spectrum taken. The transition from emission to absorption occurs over a time-scale of a couple of hours.}
        \label{fig:soar_halpha}
    \end{figure}

    The H $\alpha$ emission over our ten observations is shown in Figure~\ref{fig:soar_halpha}. Prominent double-peaked emission features are seen over the two spectra taken on May 24 2015. This emission feature is also seen in the first two spectra taken on May 28 2015, although only one peak is seen. Between 06:13 and 08:11 on May 28 2015, the emission feature disappears and is replaced with an absorption feature, indicating a change in the source over a time-scale of less than two hours. The absorption feature persists for the rest of the spectra. This behaviour is also seen in the H $\beta$, H $\gamma$, H $\delta$, and He \textsc{i} lines.
    
\subsection{Optical photometry}
    We obtained 55 minutes of optical observations with the Gemini North Acquisition Camera that were simultaneous with our radio and X-ray observations to search for variations in the V filter of NGC~6652B. The duration of each exposure was 2 s. The image subtraction software \textsc{isis} was used to extract the light curves of the source and three non-varying check sources in the field. 
    The region near the source is crowded by bright, nearby sources, and the source is located in the point-spread function of a bright source, as highlighted by the image shown in Figure~\ref{fig:images}. Stacking the snapshots to derive a deep image of the field shows that the source is not detected down to a conservative \textit{V} filter magnitude of $\sim17.7$. 
    
\section{Discussion} \label{sec:Discussion}
    We observed NGC~6652B simultaneously with \chandra and the VLA to investigate the correlation between the radio and X-ray behaviour of the source. Additional optical spectra obtained with SOAR provided a means to characterise the optical counterpart, and to search for signatures of accretion. These observations have provided several clues as to the classification of the system and the behaviours that are present. 
    
    Optical spectra obtained with SOAR have highlighted variation in the H $\alpha$ emission line, with a shift from double-peaked emission to absorption over a time scale of two hours as shown in Figure~\ref{fig:soar_halpha}. Simultaneous radio and X-ray observations place NGC~6652B on the track mostly occupied by accreting BHs on the radio--X-ray luminosity plane (Figure~\ref{fig:lrlx}), indicating that it has a higher radio luminosity than typical accreting NSs of a similar X-ray luminosity. Breaking these observations into intervals of high and low X-ray count rates hints that radio luminosity may decrease as X-ray luminosity increases and vice versa. This behaviour is only present when considering the average radio and X-ray luminosities during these regimes, and cross-correlation analysis indicates that there is no correlation between the simultaneous radio and X-ray light curves shown in Figure~\ref{fig:chandra_vla} to $>99\%$ confidence. X-ray spectral analysis of NGC~6652B indicates a hard photon index ($\Gamma\sim1.3$), resulting from non-thermal emission, and spectral fitting of the low count rate regime highlights weak evidence for intrinsic absorption. These observational properties suggest that NGC 6652B is a dim variable X-ray source, with moderately variable radio emission.
    
    NGC~6652B has been the subject of previous X-ray observations with \chandra \citep{Coomber2011, Stacey2012}. Through a \SI{5}{ks} \chandra observation, \citet{Coomber2011} observed rapid flaring variability of the source, with flares up to $L_X = \SI{9e34}{\erg\per\second}$ persisting for several minutes. The hardening of the X-ray emission during lower count rates suggested that the variability seen is caused by obscuration by an accretion disc with a high inclination angle. \citet{Stacey2012} further developed the X-ray picture of NGC~6652B with a \SI{47.5}{ks} observation of the cluster. The variable, hard emission of the source was still present, and the X-ray luminosity of the source suggested a NS primary. A tentative increase in the column density $N_{\textrm{H}}$ during low count rates was also observed, however it was concluded that this increase in column density was not sufficient enough to decrease the count rate, ruling out obscuration by the accretion disc as the cause of the source's variability. Instead, \citet{Stacey2012} suggested that the variability of the source was due to the propeller effect, where the rotating magnetosphere of the NS prevents accretion onto its surface \citep{Illarionov1975}, and classified the source as a very faint X-ray transient, due to its peak X-ray luminosity falling in the range of $10^{34-36}$ \si{\erg\per\second}.

\subsection{The companion star} \label{sec:optical_counterpart}
    The presence of the optical counterpart on the main sequence has been noted in previous studies of the source. The optical counterpart appears to fall on the main sequence in \textit{V}-\textit{I} and \textit{g}-\textit{r} colour-magnitude diagrams (CMDs) of the cluster \citep{Heinke2001,Engel2012}, but to the blue side of the main sequence in a \textit{U}-\textit{B} CMD \citep{Deutsch1998}. The position of the source on the cluster CMD also rules out the possibility of a giant companion. The 1 magnitude variability, first seen by \citep{Deutsch2000}, was identified as prohibiting the optical counterpart from actually being a main sequence star as some component of the system must be heated by X-rays, contributing a blue component to its colour \citep{Heinke2001,Engel2012}. The presence on the main sequence then suggests that the donor may be a redder object, such as a subgiant or red straggler star \citep{Engel2012}.
    
\begin{figure*}
        \centering
        \includegraphics[width=0.95\textwidth]{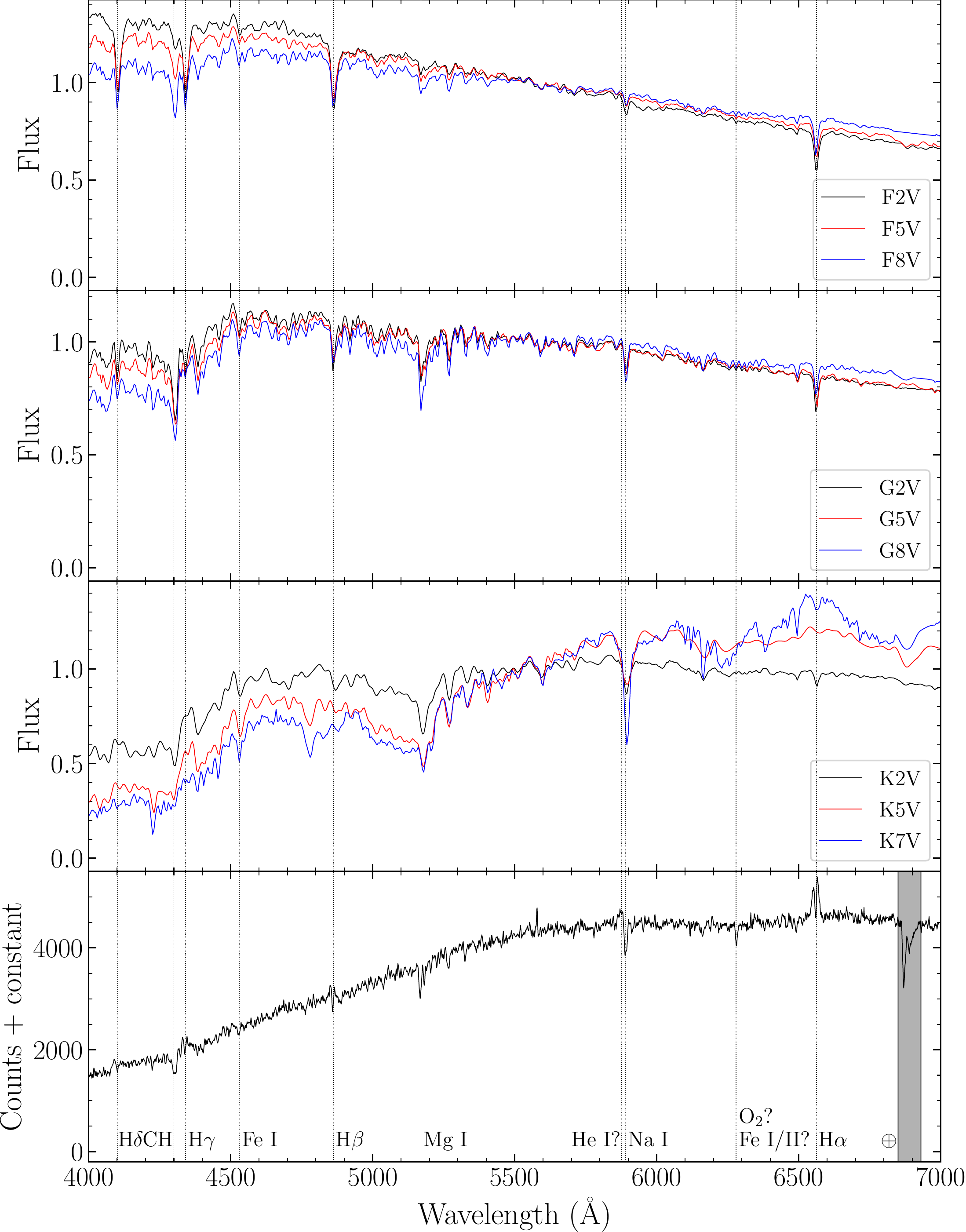}
        \caption{The stacked optical spectrum for NGC~6652B (bottom panel) compared to example stellar spectra for F, G and K-type main sequence stars. The identified spectral lines in the spectrum of NGC~6652B are indicated with black dashed lines. The spectrum of NGC~6652B best matches those of a G-type star.}
        \label{fig:star_class}
    \end{figure*}

    Based on the optical spectra shown in Figure~\ref{fig:soar_stacked}, some conclusions can be made as to the nature of the companion star in NGC~6652B. To investigate the star class of the companion, we compared the stacked optical spectrum of NGC~6652B to example stellar spectra for F, G and K-type main sequence stars from the ESO library of stellar spectra\footnote{Available from \url{https://www.eso.org/sci/facilities/paranal/decommissioned/isaac/tools/lib.html}} based on the stellar spectra catalogue produced by \citet{Pickles1998}. This comparison is shown in Figure~\ref{fig:star_class}. 
    
    From this comparison, we conclude that the most likely star class for the companion is an early to mid G-type star. The presence of the G-band means the star cannot be much cooler than a K2 star, and the other absorption features closely match what is seen in G-type stars. However, we note that we cannot be completely confident in this observational classification of the companion due to the possibility of spectral contamination. NGC~6652B is located in an optically crowded region, so it is possible that the absorption lines seen in our optical spectra could be contaminated by nearby bright sources. However, it is unlikely that there will be significant contamination of the variable H $\alpha$ emission as this would require several accreting sources within close proximity of NGC 6652B. 

    If the optical companion to NGC~6652B is a G-type star, it would be similar to that of J1023. J1023 contains a $\sim0.2\textrm{M}_{\odot}$ G-type star that is been irradiated by the pulsar in the system \citep{Archibald2013}. This source has also been observed to undergo spectral type changes from G5 to F6 depending on the side of the star that is observed \citep{Shahbaz2019}. 

\subsection{Potential source classifications for NGC~6652B} \label{sec:other_explanations}
    NGC~6652B is located within the half-light radius of the cluster, and optical spectroscopy shows that the systematic velocity of the H $\alpha$ profile is consistent with the systematic velocity of the cluster NGC~6652 and offset from the field stars in that direction, meaning that it is unlikely to be a foreground star. The zero redshift of the spectral lines shows that it is not a background AGN. We identify the star as ID number 1676 in the HUGS catalogue (their photometric method 2) of \citet{Nardiello2018}. Using proper motions between ACS and WFC3 epochs, the cluster membership probability of the star is estimated to be 97.8\%, confirming that it is a cluster member.

    There are many different types of X-ray sources within Galactic globular clusters, such as cataclysmic variables (CVs) and active binaries. CVs are prominent X-ray sources in clusters and can display X-ray variability. Three (rather extreme) examples of CVs are shown in the radio-X-ray luminosity plane in Figure~\ref{fig:lrlx}: AE Aqr, AR Sco, and SS Cyg. The brightest of these in X-rays is SS Cyg with a peak X-ray luminosity of $\sim\SI{e32}{\erg\per\second}$ \citep{Russell2016}, with the other two sources having X-ray luminosities $<\SI{e31}{\erg\per\second}$ \citep{Eracleous1991,Abada-Simon1993,Marsh2016}. NGC~6652B has an average X-ray luminosity of $\sim\SI{2e34}{\erg\per\second}$, which is two orders of magnitude higher than the observed X-ray luminosities of virtually all CVs other than intermediate polars (IPs). Furthermore, NGC~6652B has an average radio luminosity of $\sim\SI{3e28}{\erg\per\second}$, which is an order of magnitude higher than the observed radio luminosity of AR Sco \citep{Marsh2016} and two orders of magnitude higher than that of SS Cyg \citep{Russell2016}, meaning that based on radio luminosity it is unlikely that NGC~6652B is a CV.
    
    In non-core-collapsed clusters there is a single faint CV population as seen through observations of the clusters $\omega$ Centauri and 47 Tucanae \citep{Cool2013,Sandoval2018} collated by \citet{Belloni2020a}, with the brightest of this population reaching an X-ray luminosity of $\sim\SI{e33}{\erg\per\second}$. This is an order of magnitude lower than that of NGC~6652B, again meaning that based on X-ray luminosity it is unlikely that NGC~6652B is a CV. The brightest CV in the X-rays in a GC, 1E1339.8+2837 in M3, occasionally shows similar levels of X-ray brightness, variability and spectral hardness as NGC~6652B. However, that source is distinctly different from NGC~6652B in optical/UV bands, with a very bright blue UV counterpart \citep{Edmonds2004, Zhao2019}, and is not detected in the radio band \citep{Shishkovsky2020}. Another population of X-ray bright CVs are intermediate polars (IPs), which radiate most of their accretion energy as X-rays at both low and high accretion rates \citep{Patterson1985}. However, most IPs don't show substantial radio emission \citep{Barrett2017}, again making it unlikely that NGC~6652B could be classified as such.

    Active binaries are tidally locked stars in close binaries which produce X-rays due to their high rate of rotation, and make up a large portion of faint X-ray sources in clusters. These systems generally have X-ray luminosities of $<\SI{e31}{\erg\per\second}$ \citep{Gudel2002}. Additionally, the X-ray emission for coronally active stars when not flaring saturates at an X-ray luminosity of $10^{-3}$ of the bolometric luminosity \citep{Vilhu1987}. We have multiple X-ray observations of this object and have measured the average X-ray luminosity of NGC~6652B to be $\sim\SI{2e34}{\erg\per\second}$. This makes it very unlikely that NGC~6652B is an active binary.
    
    Given the position of NGC~6652B on the radio--X-ray luminosity plane in Figure~\ref{fig:lrlx}, a possible explanation for the system is that it contains an accreting compact object. NGC~6652B shows behaviours that are different to known, typical NS X-ray binaries (NSXBs), such as Centaurus X-4 (Cen X-4). Cen X-4 is the closest NSXB at a distance of $1.2\pm0.3$ \si{kpc} \citep{Chevalier1989}, and is one of the few NSXB systems that can be studied in depth to understand how these systems behave. Cen X-4 has been in quiescence for over three decades \citep{Hjellming1988}, with its X-ray luminosity ($10^{32}$ \si{\erg\per\second}) indicating low level accretion is still occurring \citep{Campana2000,Campana2004,Cackett2010,Bernardini2013}. No radio emission has been seen in quiescence down to a radio luminosity limit of $\sim10^{26}$ \si{\erg\per\second} \citep{Tudor2017a}. These observational properties are inconsistent with NGC~6652B. NGC~6652B has a 1-10 keV X-ray luminosity of $\sim\SI{1.8e34}{\erg\per\second}$, and Cen X-4 only displays similar X-ray luminosities during a rise or fall between outburst and quiescence, which are short-lived phases. The radio luminosity of NGC~6652B is also at least two orders of magnitude higher than that of Cen X-4. Due to this, it is unlikely that NGC~6652B is a ``typical'' NSXB. NGC~6652B has a higher radio luminosity than any accreting NS observed at the same X-ray luminosity and its persistent X-ray luminosity ($\sim10^{34}$) is a signature of unusual accretion. 
    
    Given that NGC~6652B displays observational properties inconsistent with typical NSXB, we can expand our comparison to more unusual classes of NS binaries, such as symbiotic X-ray binaries (SyXBs). SyXBs are a small subset of binary systems that contain a NS accreting from a late-type giant, with X-ray luminosities in the range of $10^{32-36}$ \si{\erg\per\second} \citep{Yungelson2019}. The orbital periods of SyXBs can be several years. While the X-ray luminosity of NGC~6652B falls into the range expected for SyXBs, the source cannot be a SyXB. A CMD of the cluster NGC 6652 places the optical companion to NGC~6652B on the main sequence \citep{Heinke2001,Engel2012}, and not on the giant branch as would be needed to be classified as a SyXB. Our spectral classification of the companion to NGC~6652B as a G-type star is also inconsistent with the M-type stars expected in SyXBs. Due to these factors, NGC~6652B cannot be identified as a SyXB.
    
    To test whether NGC~6652B could contain an accreting BH, we can compare this source to V404 Cygni, which is one of the most well studied BHXBs due to its proximity to Earth \citep[$2.39\pm0.14$ kpc, ][]{Miller-Jones2009} and well constrained orbital parameters \citep{Casares1992,Khargharia2010}. When the source transitioned into quiescence after its 2015 outburst, the X-ray luminosity decreased to $\sim\SI{e33}{\erg\per\second}$ and the photon index transitioned from $\Gamma\sim1.6$ to $\Gamma\sim2$ \citep{Plotkin2017}. The observed properties of NGC~6652B are inconsistent with these observations of V404 Cyg. NGC~6652B has a hard ($\Gamma\sim1.3$) X-ray spectrum which is inconsistent with quiescent BHXBs which show $\Gamma\sim2$ \citep{Plotkin2013,Reynolds2014}, making it unlikely to be a BHXB in quiescence. Further evidence against NGC~6652B containing a BH is the variable H $\alpha$ emission, unlike the broad H $\alpha$ emission BHs are expected to consistently exhibit \citep{Casares2015,Casares2016}. NGC~6652B shows double-peaked H $\alpha$ emission that transitions to complete absorption on a time-scale of a couple of hours as seen in Figure~\ref{fig:soar_halpha}, which is inconsistent with known BH properties. The cause of this variation could be an intra-binary shock or changes in the ionisation of the material in the accretion disc.

\subsection{NGC~6652B as a tMSP candidate} \label{sec:tMSP_candidate}
    Based on the analysis presented in \S~\ref{sec:Results} and the inconsistencies with several ``canonical'' source classes in \S~\ref{sec:other_explanations}, a possible explanation for the dim, variable X-ray emission and radio emission seen from NGC 6652B is that this source may be a tMSP candidate in the accreting state. We observe multiple X-ray flares in the 1-10 keV band of $>10^{34}$ \si{\erg\per\second} during the \chandra observation. The \chandra light curve in the top panel of Figure~\ref{fig:chandra_vla} shows that the source undergoes several X-ray flaring events on time-scales of 100s to 1000s of seconds, spending almost half of the observation in 1-10 keV X-ray flaring events up to $\sim$\SI{4.8e34}{\erg\per\second}. The average 1-10 keV X-ray luminosity of the source during this observation is $\sim$\SI{1.8e34}{\erg\per\second}. This is consistent with earlier observations of the source by both \citet{Coomber2011} and \citet{Stacey2012}, who reported similar unusual variability on time-scales of $\sim$\SI{100}{\second} and average X-ray luminosities of $\sim$\SI{2e34}{\erg\per\second} and \SI{1.6e34}{\erg\per\second} respectively.

    This behaviour is similar to the recently identified tMSP candidate 3FGL J0427.9-6704 \citep[J0427, ][]{Li2020}. J0427 was observed to be in the flaring mode for the entirety of a 20 hr observation, exhibiting continuous flaring up to $\sim$\SI{2e34}{\erg\per\second} on time-scales of 10s to 100s of seconds. It is possible that we are also observing NGC~6652B in the flare-dominated accretion mode of tMSPs due to the magnitude and frequency of the flaring events over the course of the X-ray observation. The long-duration flaring behaviour of NGC~6652B is also similar to that of known tMSPs, such as  XSS J12270-4859 (J12270) which has been shown to display several flares on time-scales of 100s of seconds during its accretion-dominated phase \citep{deMartino2010,deMartino2013}, and J1023, which displayed a long flaring mode of duration $\sim$10 hr after its state transition in 2013 \citep{Tendulkar2014}. 
    
    NGC~6652B also displays similarities to the new tMSP candidate 4FGL J0540.0-7552 \citep{Strader2021}. This source displays a hard X-ray spectrum, and X-ray and optical variability. Additionally, the source appears to consistently be in an X-ray flaring mode, similar to the behaviour we observe in NGC~6652B and reported for J0427. This indicates that persistent flaring activity could be a common property in tMSPs.

    The radio continuum behaviour of NGC~6652B is also consistent with that of J0427. NGC~6652B displays suggestive evidence for variability throughout the duration of the VLA observation, and J0427 shows some variation in radio emission with no flaring events \citep{Li2020}. This variation is also seen on longer time-scales, with previous observations of NGC~6652B from 2016 also showing similar moderate variability. Both sources show no evidence that the radio emission is eclipsed, which is consistent with steady radio outflow.
 
    When we split the radio and X-ray light curves of NGC~6652B by X-ray count rate and split them into regimes of high (>\SI{0.09}{counts\per\second}) and low ($\leq0.09$ \si{counts\per\second}) X-ray counts, there is a hint that radio luminosity may decrease as X-ray luminosity increases and vice versa. On the radio--X-ray luminosity plane (Figure~\ref{fig:lrlx}), NGC~6652B sits on the track accreting BHs occupy, and the source displays the same behaviour as J1023, where its position on this plot changes depending on what X-ray count rate regime it is in, allowing it to move above or below the BH correlation. However, it should be noted that this parallel with J1023 is not exact. The position of J1023 changes between the low and high mode, whereas it is unknown whether NGC~6652B has any X-ray modes other than the flaring mode. J0427 also changes position on the radio--X-ray luminosity plane, and sits on the accreting BH track. Both NGC~6652B and J0427 display a radio/X-ray luminosity ratio much higher than any known tMSP, however J0427 displays a higher radio luminosity with higher X-ray luminosities and instead undergoes marginal shifts along the BH correlation during changes in luminosity.
    
    X-ray spectral analysis (\S~\ref{sec:xray_spec}) indicates that there is some evidence for changes in the hardness of the emission and column density during different X-ray count rate regimes. For low counts ($\leq0.09$ counts/s) the X-ray emission is harder, while it is slightly softer emission at higher count rates. These properties have been previously observed, with both \cite{Coomber2011} and \citet{Stacey2012} observing a tendency for harder emission to be present in low count rate regimes. Additionally, there is marginal evidence that the column density towards the source changes subtly based on the X-ray count rate regime (intrinsic $N_H$ in Table~\ref{tab:x-ray_spec}). While this inconsistency is only seen to a $\geq1\sigma$ level as seen in the contour plots shown in Figure~\ref{fig:x-ray_contour}, this behaviour of NGC~6652B has also been previously noted by \citet{Coomber2011} and \citet{Stacey2012}, with \citet{Stacey2012} concluding that these changes were due to an instability involving the propeller effect \citep{Illarionov1975}. Furthermore, this conclusion was made before the confirmation of the existence of tMSPs, and since their discovery it has been discussed that the propeller effect may be present in these systems \citep{Papitto2015,Deller2015}. This suggests that we may be observing NGC~6652B in an accreting state similar to that of tMSPs where the propeller effect is present. The changes in column density and the presence of intrinsic absorption would then come from the propeller-driven outflow. This provides a more physical basis for identifying NGC~6652B as a possible tMSP candidate.
    
    Gamma-ray emission has been detected from NGC 6652 \citep{TheFermiLATcollaboration2010}, which would be consistent with a tMSP being present within the cluster. Accreting tMSPs are known to be gamma-ray sources, with the gamma-ray flux of these systems increasing by a factor of $\sim5$ during a state change to the accreting state \citep{Stappers2014,Deller2015}. MSPs are also gamma-ray emitters, and NGC 6652 contains one known MSP \citep{DeCesar2011,DeCesar2015} which is not associated with NGC~6652B. It is unclear whether the gamma-ray emission originates from one of these two sources or some other unidentified source, and further investigation into this emission is needed. 

    NGC~6652B also displays double-peaked H $\alpha$ emission, which is a signature for accreting binary systems \citep[e.g.,][]{Casares2015,Casares2016}. However, the H $\alpha$ emission is variable, and goes from double-peaked emission to full absorption over a time-scale of a couple of hours. This is similar to the behaviour of the NS binary J1417.7-4402 \citep[J1417,][]{Swihart2018}. J1417 displays H $\alpha$ profiles that vary in amplitude and shape over the course of the 5.4 day orbital period of the binary system, and it was thought that these variations could be explained by the interactions between the pulsar wind, and a magnetically driven wind from the companion star. If such an intra-binary shock is present in NGC~6652B, it would disfavour the presence of an accretion disc. However, J1417 is a different kind of system to NGC~6652B (see \S \ref{sec:optical_counterpart}), consisting of a giant companion in a wide orbit, and may display behaviours different to those present in NGC~6652B. These variations could also be caused by changes in the accretion disc. As the material in the disc becomes ionised, the emission features would disappear, potentially explaining the rapid change from emission to absorption seen in the H $\alpha$ profile of NGC~6652B.

    While NGC~6652B displays many properties consistent with known tMSPs and tMSP candidates, there are some inconsistencies with this classification. A feature distinct to tMSPs, the bimodal X-ray flux distribution, is not seen in the X-ray light curves of NGC~6652B. The lack of a bimodal X-ray flux distribution may be due to a combination of the light curve being dominated by flares and/or the effect of the relatively large distance.
    
    Another feature distinct to tMSPs is the anti-correlation between the radio and X-ray emission between the ``active'' and ``passive'' modes. As shown by \citet{Bogdanov2018} through simultaneous \chandra and VLA observations of J1023, a transition into the X-ray low mode is accompanied with a corresponding increase in radio flux density, with the radio flux density decreasing as the X-ray count rate increases on the transition back to the high mode. A consequence of this behaviour is that during the X-ray low mode, tMSPs can enter parts of the radio--X-ray luminosity plane that were thought to be exclusive to accreting BHs. 
    
    We do not observe any clear evidence of anti-correlated behaviour between the radio and X-ray light curves of NGC~6652B (Figure~\ref{fig:chandra_vla}). For a robust characterisation of the anti-correlation in and to test this method on tMSPs, we calculated the ZDCF for J1023, shown in the bottom panel of Figure~\ref{fig:cross-correlation}, based on the radio and X-ray light curves presented in \citet{Bogdanov2018}. The radio light curve is binned at 30 s, and the X-ray light curve is binned at 25 s. These light curves show an X-ray flare and several radio flares, so we investigated how the correlation function behaves when these flaring events are included in and removed from the datasets. The flares are those identified in \citet{Bogdanov2018}. With the flares removed from the datasets, we see a strong anti-correlation between the two light curves at a time lag of $\sim 0$ hr. However, it is clear that the flares in tMSPs (which do not show any anti-correlation between the X-rays and radio data) can affect the correlation function, to a level where significant anti-correlation in a light curve dominated by flares could vanish entirely. When we calculate the ZDCF for J1023 using the full datasets (including the flares), the anti-correlation disappears (Figure~\ref{fig:cross-correlation}). We do not see any clear anti-correlation in the correlation function of NGC~6652B to $>99\%$. A possible reason for this is that NGC~6652B is much further away than J1023, such that we do not have sufficient sensitivity to differentiate between the canonical high and low modes. We also observe NGC~6652B in a largely X-ray flare dominated state, which may also explain why we don't observe any significant anti-correlation between any existing X-ray high and low mode, again lending credence that NGC~6652B, J0427, and 4FGL J0540.0-7552, can be considered ``flare-type'' tMSPs.

\section{Conclusions and future work}
    We conclude based on simultaneous \chandra and VLA observations in addition to older optical spectra that NGC~6652B is best explained as a tMSP candidate in the accreting state. NGC~6652B displays similar behaviour to other tMSPs and tMSP candidates, most prominently several X-ray flares observed over the course of the \chandra observation. This behaviour is similar to the flaring behaviour of the tMSPs PSR J1023+0038 and XSS J12270-4859, and the new tMSP candidates 3FGL J0427.9-6704 \citep{Li2020} and 4FGL J0540.0-7552 \citep{Strader2021}. This prolonged flaring behaviour suggests that we are observing NGC 6652B in a flare-dominated mode which is consistent with that observed from 3FGL J0427.9-6704 and 4FGL J0540.0-7552. The presence of doubled-peaked H $\alpha$ emission in previous optical spectra is indicative of the presence of an accretion disc in the system at one point in the past five years, further supporting the conclusion that we are observing NGC~6652B in the accreting state. However, we observe a transition to complete H $\alpha$ absorption, and an alternate explanation to this is that the H $\alpha$ profile in this system is caused by the shocks between the pulsar wind and the wind from the companion star, similar to those observed in J1417.7-4402 \citep{Swihart2018}.
    
    Other scenarios as to the nature of NGC~6652B include a CV, an active binary, NSXB, and a BHXB; however each of these scenarios are considered less likely than a tMSP candidate explanation. NGC~6652B has a higher radio and X-ray luminosity than known CVs and a higher X-ray luminosity than known active binaries. NGC~6652B has a higher radio luminosity than NSs with similar X-ray luminosities, and while NGC~6652B sits near the correlation for accreting BHs, it is unlikely to be a BHXB as it displays a hard X-ray spectrum.
    
    An important caveat to the classification of NGC~6652B as a tMSP candidate is that we cannot confirm that it is a tMSP until a state change is observed. Further observations of the source are needed. In particular, NGC~6652B would need to be monitored for any transition into the rotation-powered MSP state, which would confirm its transitional nature. Additional simultaneous radio and X-ray observations would allow for any correlation between the radio and X-ray emission to be probed, and also indicate whether the source is still in a mode dominated by X-ray flares. Further optical spectroscopy with a larger (8 m) telescope would allow for a more detailed study of the optical counterpart. However, identifying new tMSP candidates such as NGC~6652B is still of great importance to tMSP science as it highlights the range of interesting multi-wavelength behaviours that these sources exhibit, and again demonstrates the usefulness of simultaneous radio and X-ray observations for identifying known observational characteristics of tMSPs.

\section*{Acknowledgements}
    We thank the anonymous referee for feedback on this manuscript. AP and AB thank Alexandra Tetarenko for helpful discussions. AP was supported by an Australian Government Research Training Program (RTP) Stipend and RTP Fee-Offset Scholarship through Federation University Australia. JCAM-J is the recipient of an Australian Research Council Future Fellowship (FT140101082). JS acknowledges support from NSF grants AST-1308124 and AST-1714825 and from the Packard Foundation. COH thanks Andrew Stephens, Morten Andersen, and the Gemini staff for assistance with our Gemini DDT observations. COH is supported in part by NSERC Discovery Grant RGPIN-2016-04602. The Australia Telescope Compact Array is part of the Australia Telescope National Facility which is funded by the Australian Government for operation as a National Facility managed by CSIRO. We acknowledge the Gomeroi people as the traditional owners of the Observatory site. The National Radio Astronomy Observatory is a facility of the National Science Foundation operated under cooperative agreement by Associated Universities, Inc.. We acknowledge the use of the following packages/software in this work: \textsc{casa} \citep{McMullin2007}, provided by the National Radio Astronomy Observatory (NRAO); \textsc{ciao} \citep{Fruscione2006}, provided by the \chandra X-ray centre (CXC); \textsc{heasoft}, provided by the High Energy Astrophysics Science Archive Research Centre (HEASARC); \textsc{isis} \citep{Alard1998,Alard2000}; the \textsc{fortran} implementation of the ZDCF algorithm by \citet{Alexander1997}; and the \textsc{python} packages \textsc{astropy} \citep{AstropyCollaboration2018}, \textsc{ccdproc} \citep{ccdproc}, \textsc{matplotlib} \citep{Hunter2007}, and \textsc{numpy} \citep{numpy}. This work made use of NASA's Astrophysics Data System and arXiv.

\section*{Data Availability}
    The data underlying this work are available in the following locations: X-ray data is available from the \chandra Data Archive at \url{https://cda.harvard.edu/chaser/} under the observation ID 18987; radio data is available from the NRAO Science Data Archive at \url{https://archive.nrao.edu/archive/} under the project codes SI0399 and 16A-325; optical photometric data is available from the Gemini Observatory Archive at \url{https://archive.gemini.edu/searchform} under the program ID GN-2017A-DD-7; and optical spectroscopic data at the NOAO archive at \url{http://archive1.dm.noao.edu/}.




\bibliographystyle{mnras}
\bibliography{all_references} 

\begin{thebibliography}{}
\makeatletter
\relax
\def\mn@urlcharsother{\let\do\@makeother \do\$\do\&\do\#\do\^\do\_\do\%\do\~}
\def\mn@doi{\begingroup\mn@urlcharsother \@ifnextchar [ {\mn@doi@}
  {\mn@doi@[]}}
\def\mn@doi@[#1]#2{\def\@tempa{#1}\ifx\@tempa\@empty \href
  {http://dx.doi.org/#2} {doi:#2}\else \href {http://dx.doi.org/#2} {#1}\fi
  \endgroup}
\def\mn@eprint#1#2{\mn@eprint@#1:#2::\@nil}
\def\mn@eprint@arXiv#1{\href {http://arxiv.org/abs/#1} {{\tt arXiv:#1}}}
\def\mn@eprint@dblp#1{\href {http://dblp.uni-trier.de/rec/bibtex/#1.xml}
  {dblp:#1}}
\def\mn@eprint@#1:#2:#3:#4\@nil{\def\@tempa {#1}\def\@tempb {#2}\def\@tempc
  {#3}\ifx \@tempc \@empty \let \@tempc \@tempb \let \@tempb \@tempa \fi \ifx
  \@tempb \@empty \def\@tempb {arXiv}\fi \@ifundefined
  {mn@eprint@\@tempb}{\@tempb:\@tempc}{\expandafter \expandafter \csname
  mn@eprint@\@tempb\endcsname \expandafter{\@tempc}}}

\bibitem[\protect\citeauthoryear{{Abada-Simon}, Lecacheux, Bastian, Bookbinder
  \& Dulk}{{Abada-Simon} et~al.}{1993}]{Abada-Simon1993}
{Abada-Simon} M.,  Lecacheux A.,  Bastian T.~S.,  Bookbinder J.~A.,   Dulk
  G.~A.,  1993, \mn@doi [The Astrophysical Journal] {10.1086/172479}, 406, 692

\bibitem[\protect\citeauthoryear{Alard}{Alard}{2000}]{Alard2000}
Alard C.,  2000, \mn@doi [Astronomy and Astrophysics Supplement Series]
  {10.1051/aas:2000214}, 144, 363

\bibitem[\protect\citeauthoryear{Alard \& Lupton}{Alard \&
  Lupton}{1998}]{Alard1998}
Alard C.,  Lupton R.~H.,  1998, \mn@doi [The Astrophysical Journal]
  {10.1086/305984}, 503, 325

\bibitem[\protect\citeauthoryear{Alexander}{Alexander}{1997}]{Alexander1997}
Alexander T.,  1997, in Maoz D.,  Sternberg A.,   Leibowitz E.~M.,  eds,
  Astrophysics and {{Space Science Library}}, Vol.~218, Astronomical {{Time
  Series}}.
p.~163, \mn@doi{10.1007/978-94-015-8941-3_14}

\bibitem[\protect\citeauthoryear{Alpar, Cheng, Ruderman  \& Shaham}{Alpar
  et~al.}{1982}]{Alpar1982}
Alpar M.~A.,  Cheng A.~F.,  Ruderman M.~A.,   Shaham J.,  1982, \mn@doi
  [Nature] {10.1038/300728a0}, 300, 728

\bibitem[\protect\citeauthoryear{Ambrosino et~al.}{Ambrosino
  et~al.}{2017}]{Ambrosino2017}
Ambrosino F.,  et~al., 2017, \mn@doi [Nature Astronomy (2017), published
  on-line on October 2, 2017] {10.1038/s41550-017-0266-2}, 1, 854

\bibitem[\protect\citeauthoryear{Archibald et~al.}{Archibald
  et~al.}{2009}]{Archibald2009}
Archibald A.~M.,  et~al., 2009, \mn@doi [Science] {10.1126/science.1172740},
  324, 1411

\bibitem[\protect\citeauthoryear{Archibald, Kaspi, Hessels, Stappers, Janssen
  \& Lyne}{Archibald et~al.}{2013}]{Archibald2013}
Archibald A.~M.,  Kaspi V.~M.,  Hessels J. W.~T.,  Stappers B.,  Janssen G.,
  Lyne A.,  2013, arXiv e-prints, 1311, arXiv:1311.5161

\bibitem[\protect\citeauthoryear{{Arnaud}}{{Arnaud}}{1996}]{Arnaud1996}
{Arnaud} K.~A.,  1996, in {Jacoby} G.~H.,  {Barnes} J.,  eds,  Astronomical
  Society of the Pacific Conference Series Vol. 101, Astronomical Data Analysis
  Software and Systems V. p.~17

\bibitem[\protect\citeauthoryear{{Astropy Collaboration} et~al.,}{{Astropy
  Collaboration} et~al.}{2018}]{AstropyCollaboration2018}
{Astropy Collaboration} et~al., 2018, \mn@doi [\apj]
  {10.3847/1538-3881/aabc4f}, 156, 123

\bibitem[\protect\citeauthoryear{Bahramian, Heinke, Degenaar, Chomiuk,
  Wijnands, Strader, Ho  \& Pooley}{Bahramian et~al.}{2015}]{Bahramian2015}
Bahramian A.,  Heinke C.~O.,  Degenaar N.,  Chomiuk L.,  Wijnands R.,  Strader
  J.,  Ho W. C.~G.,   Pooley D.,  2015, \mn@doi [Monthly Notices of the Royal
  Astronomical Society] {10.1093/mnras/stv1585}, 452, 3475

\bibitem[\protect\citeauthoryear{Bahramian et~al.,}{Bahramian
  et~al.}{2018a}]{Bahramian2018}
Bahramian A.,  et~al., 2018a, Radio/{{X}}-Ray Correlation Database for
  {{X}}-Ray Binaries, \mn@doi{10.5281/zenodo.1252036.
}

\bibitem[\protect\citeauthoryear{Bahramian et~al.,}{Bahramian
  et~al.}{2018b}]{Bahramian2018a}
Bahramian A.,  et~al., 2018b, \mn@doi [The Astrophysical Journal]
  {10.3847/1538-4357/aad68b}, 864, 28

\bibitem[\protect\citeauthoryear{Bahramian et~al.,}{Bahramian
  et~al.}{2020}]{Bahramian2020}
Bahramian A.,  et~al., 2020, \mn@doi [The Astrophysical Journal]
  {10.3847/1538-4357/aba51d}, 901, 57

\bibitem[\protect\citeauthoryear{{Barrett}, {Dieck}, {Beasley}, {Singh}  \&
  {Mason}}{{Barrett} et~al.}{2017}]{Barrett2017}
{Barrett} P.~E.,  {Dieck} C.,  {Beasley} A.~J.,  {Singh} K.~P.,   {Mason}
  P.~A.,  2017, \mn@doi [\aj] {10.3847/1538-3881/aa93ff}, \href
  {https://ui.adsabs.harvard.edu/abs/2017AJ....154..252B} {154, 252}

\bibitem[\protect\citeauthoryear{Bassa et~al.}{Bassa et~al.}{2014}]{Bassa2014}
Bassa C.~G.,  et~al., 2014, \mn@doi [Mon. Not. Roy. Astron. Soc.]
  {10.1093/mnras/stu708}, 441, 1825

\bibitem[\protect\citeauthoryear{Baumgardt, Hilker, Sollima  \&
  Bellini}{Baumgardt et~al.}{2019}]{Baumgardt2019}
Baumgardt H.,  Hilker M.,  Sollima A.,   Bellini A.,  2019, VizieR Online Data
  Catalog, p. J/MNRAS/482/5138

\bibitem[\protect\citeauthoryear{Belloni \& Rivera~Sandoval}{Belloni \&
  Rivera~Sandoval}{2020}]{Belloni2020a}
Belloni D.,  Rivera~Sandoval L.~E.,  2020, arXiv e-prints, 2008,
  arXiv:2008.12772

\bibitem[\protect\citeauthoryear{{Bernardini}, {Cackett}, {Brown}, {D'Angelo},
  {Degenaar}, {Miller}, {Reynolds}  \& {Wijnands}}{{Bernardini}
  et~al.}{2013}]{Bernardini2013}
{Bernardini} F.,  {Cackett} E.~M.,  {Brown} E.~F.,  {D'Angelo} C.,  {Degenaar}
  N.,  {Miller} J.~M.,  {Reynolds} M.,   {Wijnands} R.,  2013, \mn@doi [\mnras]
  {10.1093/mnras/stt1741}, \href
  {https://ui.adsabs.harvard.edu/abs/2013MNRAS.436.2465B} {436, 2465}

\bibitem[\protect\citeauthoryear{Bogdanov \& Halpern}{Bogdanov \&
  Halpern}{2015}]{Bogdanov2015}
Bogdanov S.,  Halpern J.~P.,  2015, \mn@doi [Astrophys. J.]
  {10.1088/2041-8205/803/2/L27}, 803, L27

\bibitem[\protect\citeauthoryear{Bogdanov et~al.}{Bogdanov
  et~al.}{2015}]{Bogdanov2015a}
Bogdanov S.,  et~al., 2015, \mn@doi [Astrophys. J.]
  {10.1088/0004-637X/806/2/148}, 806, 148

\bibitem[\protect\citeauthoryear{Bogdanov et~al.,}{Bogdanov
  et~al.}{2018}]{Bogdanov2018}
Bogdanov S.,  et~al., 2018, \mn@doi [Astrophys. J.] {10.3847/1538-4357/aaaeb9},
  856, 54

\bibitem[\protect\citeauthoryear{{Cackett}, {Brown}, {Miller}  \&
  {Wijnands}}{{Cackett} et~al.}{2010}]{Cackett2010}
{Cackett} E.~M.,  {Brown} E.~F.,  {Miller} J.~M.,   {Wijnands} R.,  2010,
  \mn@doi [\apj] {10.1088/0004-637X/720/2/1325}, \href
  {https://ui.adsabs.harvard.edu/abs/2010ApJ...720.1325C} {720, 1325}

\bibitem[\protect\citeauthoryear{{Campana}, {Stella}, {Mereghetti}  \&
  {Cremonesi}}{{Campana} et~al.}{2000}]{Campana2000}
{Campana} S.,  {Stella} L.,  {Mereghetti} S.,   {Cremonesi} D.,  2000, \aap,
  \href {https://ui.adsabs.harvard.edu/abs/2000A&A...358..583C} {358, 583}

\bibitem[\protect\citeauthoryear{{Campana}, {Israel}, {Stella}, {Gastaldello}
  \& {Mereghetti}}{{Campana} et~al.}{2004}]{Campana2004}
{Campana} S.,  {Israel} G.~L.,  {Stella} L.,  {Gastaldello} F.,   {Mereghetti}
  S.,  2004, \mn@doi [\apj] {10.1086/380194}, \href
  {https://ui.adsabs.harvard.edu/abs/2004ApJ...601..474C} {601, 474}

\bibitem[\protect\citeauthoryear{Casares}{Casares}{2015}]{Casares2015}
Casares J.,  2015, \mn@doi [The Astrophysical Journal]
  {10.1088/0004-637X/808/1/80}, 808, 80

\bibitem[\protect\citeauthoryear{Casares}{Casares}{2016}]{Casares2016}
Casares J.,  2016, \mn@doi [The Astrophysical Journal]
  {10.3847/0004-637X/822/2/99}, 822, 99

\bibitem[\protect\citeauthoryear{{Casares}, {Charles}  \& {Naylor}}{{Casares}
  et~al.}{1992}]{Casares1992}
{Casares} J.,  {Charles} P.~A.,   {Naylor} T.,  1992, \mn@doi [\nat]
  {10.1038/355614a0}, \href
  {https://ui.adsabs.harvard.edu/abs/1992Natur.355..614C} {355, 614}

\bibitem[\protect\citeauthoryear{{Chevalier}, {Ilovaisky}, {van Paradijs},
  {Pedersen}  \& {van der Klis}}{{Chevalier} et~al.}{1989}]{Chevalier1989}
{Chevalier} C.,  {Ilovaisky} S.~A.,  {van Paradijs} J.,  {Pedersen} H.,   {van
  der Klis} M.,  1989, \aap, \href
  {https://ui.adsabs.harvard.edu/abs/1989A&A...210..114C} {210, 114}

\bibitem[\protect\citeauthoryear{Cool, Haggard, Arias, Brochmann, Dorfman,
  Gafford, White  \& Anderson}{Cool et~al.}{2013}]{Cool2013}
Cool A.~M.,  Haggard D.,  Arias T.,  Brochmann M.,  Dorfman J.,  Gafford A.,
  White V.,   Anderson J.,  2013, \mn@doi [The Astrophysical Journal]
  {10.1088/0004-637X/763/2/126}, 763, 126

\bibitem[\protect\citeauthoryear{Coomber, Heinke, Cohn, Lugger  \&
  Grindlay}{Coomber et~al.}{2011}]{Coomber2011}
Coomber G.,  Heinke C.~O.,  Cohn H.~N.,  Lugger P.~M.,   Grindlay J.~E.,  2011,
  \mn@doi [The Astrophysical Journal] {10.1088/0004-637X/735/2/95}, 735, 95

\bibitem[\protect\citeauthoryear{{Coti Zelati} et~al.,}{{Coti Zelati}
  et~al.}{2019}]{CotiZelati2019}
{Coti Zelati} F.,  et~al., 2019, \mn@doi [\aap] {10.1051/0004-6361/201834835},
  \href {https://ui.adsabs.harvard.edu/abs/2019A&A...622A.211C} {622, A211}

\bibitem[\protect\citeauthoryear{Craig et~al.,}{Craig et~al.}{2017}]{ccdproc}
Craig M.,  et~al., 2017, astropy/ccdproc: v1.3.0.post1,
  \mn@doi{10.5281/zenodo.1069648}, \url
  {https://doi.org/10.5281/zenodo.1069648}

\bibitem[\protect\citeauthoryear{DeCesar, Ransom  \& Ray}{DeCesar
  et~al.}{2011}]{DeCesar2011}
DeCesar M.~E.,  Ransom S.~M.,   Ray P.~S.,  2011, arXiv:1111.0365 [astro-ph]

\bibitem[\protect\citeauthoryear{DeCesar, Ransom, Kaplan, Ray  \&
  Geller}{DeCesar et~al.}{2015}]{DeCesar2015}
DeCesar M.~E.,  Ransom S.~M.,  Kaplan D.~L.,  Ray P.~S.,   Geller A.~M.,  2015,
  \mn@doi [The Astrophysical Journal Letters] {10.1088/2041-8205/807/2/L23},
  807, L23

\bibitem[\protect\citeauthoryear{Deller et~al.,}{Deller
  et~al.}{2015}]{Deller2015}
Deller A.~T.,  et~al., 2015, \mn@doi [Astrophys. J.]
  {10.1088/0004-637X/809/1/13}, 809, 13

\bibitem[\protect\citeauthoryear{Deutsch, Margon  \& Anderson}{Deutsch
  et~al.}{1998}]{Deutsch1998}
Deutsch E.~W.,  Margon B.,   Anderson S.~F.,  1998, \mn@doi [The Astronomical
  Journal] {10.1086/300512}, 116, 1301

\bibitem[\protect\citeauthoryear{Deutsch, Margon  \& Anderson}{Deutsch
  et~al.}{2000}]{Deutsch2000}
Deutsch E.~W.,  Margon B.,   Anderson S.~F.,  2000, \mn@doi [The Astrophysical
  Journal Letters] {10.1086/312486}, 530, L21

\bibitem[\protect\citeauthoryear{{Edmonds}, {Kahabka}  \& {Heinke}}{{Edmonds}
  et~al.}{2004}]{Edmonds2004}
{Edmonds} P.~D.,  {Kahabka} P.,   {Heinke} C.~O.,  2004, \mn@doi [\apj]
  {10.1086/422136}, \href
  {https://ui.adsabs.harvard.edu/abs/2004ApJ...611..413E} {611, 413}

\bibitem[\protect\citeauthoryear{Engel, Heinke, Sivakoff, Elshamouty  \&
  Edmonds}{Engel et~al.}{2012}]{Engel2012}
Engel M.~C.,  Heinke C.~O.,  Sivakoff G.~R.,  Elshamouty K.~G.,   Edmonds
  P.~D.,  2012, \mn@doi [The Astrophysical Journal]
  {10.1088/0004-637X/747/2/119}, 747, 119

\bibitem[\protect\citeauthoryear{Eracleous, Halpern  \& Patterson}{Eracleous
  et~al.}{1991}]{Eracleous1991}
Eracleous M.,  Halpern J.,   Patterson J.,  1991, \mn@doi [The Astrophysical
  Journal] {10.1086/170716}, 382, 290

\bibitem[\protect\citeauthoryear{{Fender} \& {Kuulkers}}{{Fender} \&
  {Kuulkers}}{2001}]{Fender2001}
{Fender} R.~P.,  {Kuulkers} E.,  2001, \mn@doi [\mnras]
  {10.1046/j.1365-8711.2001.04345.x}, \href
  {https://ui.adsabs.harvard.edu/abs/2001MNRAS.324..923F} {324, 923}

\bibitem[\protect\citeauthoryear{{Fender}, {Gallo}  \& {Jonker}}{{Fender}
  et~al.}{2003}]{Fender2003}
{Fender} R.~P.,  {Gallo} E.,   {Jonker} P.~G.,  2003, \mn@doi [\mnras]
  {10.1046/j.1365-8711.2003.06950.x}, \href
  {https://ui.adsabs.harvard.edu/abs/2003MNRAS.343L..99F} {343, L99}

\bibitem[\protect\citeauthoryear{{Foight}, {G{\"u}ver}, {{\"O}zel}  \&
  {Slane}}{{Foight} et~al.}{2016}]{Foight2016}
{Foight} D.~R.,  {G{\"u}ver} T.,  {{\"O}zel} F.,   {Slane} P.~O.,  2016,
  \mn@doi [\apj] {10.3847/0004-637X/826/1/66}, \href
  {https://ui.adsabs.harvard.edu/abs/2016ApJ...826...66F} {826, 66}

\bibitem[\protect\citeauthoryear{Fruscione et~al.,}{Fruscione
  et~al.}{2006}]{Fruscione2006}
Fruscione A.,  et~al., 2006, in \textbackslash procspie. p. 62701V,
  \mn@doi{10.1117/12.671760}

\bibitem[\protect\citeauthoryear{Gallo, Fender  \& Pooley}{Gallo
  et~al.}{2003}]{Gallo2003}
Gallo E.,  Fender R.~P.,   Pooley G.~G.,  2003, \mn@doi [Mon. Not. Roy. Astron.
  Soc.] {10.1046/j.1365-8711.2003.06791.x}, 344, 60

\bibitem[\protect\citeauthoryear{Gallo et~al.,}{Gallo et~al.}{2014}]{Gallo2014}
Gallo E.,  et~al., 2014, \mn@doi [\mnras] {10.1093/mnras/stu1599}, 445, 290

\bibitem[\protect\citeauthoryear{G{\"u}del}{G{\"u}del}{2002}]{Gudel2002}
G{\"u}del M.,  2002, \mn@doi [\araa] {10.1146/annurev.astro.40.060401.093806},
  40, 217

\bibitem[\protect\citeauthoryear{Harris}{Harris}{1996}]{Harris1996}
Harris W.~E.,  1996, \mn@doi [\apj] {10.1086/118116}, 112, 1487

\bibitem[\protect\citeauthoryear{Harris et~al.,}{Harris et~al.}{2020}]{numpy}
Harris C.~R.,  et~al., 2020, \mn@doi [Nature] {10.1038/s41586-020-2649-2}, 585,
  357–362

\bibitem[\protect\citeauthoryear{Heinke, Edmonds  \& Grindlay}{Heinke
  et~al.}{2001}]{Heinke2001}
Heinke C.~O.,  Edmonds P.~D.,   Grindlay J.~E.,  2001, \mn@doi [The
  Astrophysical Journal] {10.1086/323493}, 562, 363

\bibitem[\protect\citeauthoryear{Heinke, Bahramian, Degenaar  \& {Wijnand
  s}}{Heinke et~al.}{2015}]{Heinke2015}
Heinke C.~O.,  Bahramian A.,  Degenaar N.,   {Wijnand s} R.,  2015, \mn@doi
  [\textbackslash mnras] {10.1093/mnras/stu2652}, 447, 3034

\bibitem[\protect\citeauthoryear{{Hill} et~al.,}{{Hill}
  et~al.}{2011}]{Hill2011}
{Hill} A.~B.,  et~al., 2011, \mn@doi [\mnras]
  {10.1111/j.1365-2966.2011.18692.x}, \href
  {https://ui.adsabs.harvard.edu/abs/2011MNRAS.415..235H} {415, 235}

\bibitem[\protect\citeauthoryear{{Hjellming}, {Calovini}, {Han}  \&
  {Cordova}}{{Hjellming} et~al.}{1988}]{Hjellming1988}
{Hjellming} R.~M.,  {Calovini} T.~A.,  {Han} X.~H.,   {Cordova} F.~A.,  1988,
  \mn@doi [\apjl] {10.1086/185343}, \href
  {https://ui.adsabs.harvard.edu/abs/1988ApJ...335L..75H} {335, L75}

\bibitem[\protect\citeauthoryear{{Hunter}}{{Hunter}}{2007}]{Hunter2007}
{Hunter} J.~D.,  2007, \mn@doi [Computing in Science and Engineering]
  {10.1109/MCSE.2007.55}, \href
  {https://ui.adsabs.harvard.edu/abs/2007CSE.....9...90H} {9, 90}

\bibitem[\protect\citeauthoryear{Illarionov \& Sunyaev}{Illarionov \&
  Sunyaev}{1975}]{Illarionov1975}
Illarionov A.~F.,  Sunyaev R.~A.,  1975, Astronomy and Astrophysics, 39, 185

\bibitem[\protect\citeauthoryear{{Johnson} et~al.,}{{Johnson}
  et~al.}{2015}]{Johnson2015}
{Johnson} T.~J.,  et~al., 2015, \mn@doi [\apj] {10.1088/0004-637X/806/1/91},
  \href {https://ui.adsabs.harvard.edu/abs/2015ApJ...806...91J} {806, 91}

\bibitem[\protect\citeauthoryear{{Khargharia}, {Froning}  \&
  {Robinson}}{{Khargharia} et~al.}{2010}]{Khargharia2010}
{Khargharia} J.,  {Froning} C.~S.,   {Robinson} E.~L.,  2010, \mn@doi [\apj]
  {10.1088/0004-637X/716/2/1105}, \href
  {https://ui.adsabs.harvard.edu/abs/2010ApJ...716.1105K} {716, 1105}

\bibitem[\protect\citeauthoryear{Li, Strader, {Miller-Jones}, Heinke  \&
  Chomiuk}{Li et~al.}{2020}]{Li2020}
Li K.-L.,  Strader J.,  {Miller-Jones} J. C.~A.,  Heinke C.~O.,   Chomiuk L.,
  2020, \mn@doi [The Astrophysical Journal] {10.3847/1538-4357/ab8f28}, 895, 89

\bibitem[\protect\citeauthoryear{Linares}{Linares}{2014}]{Linares2014}
Linares M.,  2014, \mn@doi [The Astrophysical Journal]
  {10.1088/0004-637X/795/1/72}, 795, 72

\bibitem[\protect\citeauthoryear{{Maccarone}}{{Maccarone}}{2005}]{Maccarone2005}
{Maccarone} T.~J.,  2005, \mn@doi [\mnras] {10.1111/j.1745-3933.2005.00047.x},
  \href {https://ui.adsabs.harvard.edu/abs/2005MNRAS.360L..68M} {360, L68}

\bibitem[\protect\citeauthoryear{Marsh et~al.,}{Marsh et~al.}{2016}]{Marsh2016}
Marsh T.~R.,  et~al., 2016, \mn@doi [Nature] {10.1038/nature18620}, 537, 374

\bibitem[\protect\citeauthoryear{McMullin, Waters, Schiebel, Young  \&
  Golap}{McMullin et~al.}{2007}]{McMullin2007}
McMullin J.~P.,  Waters B.,  Schiebel D.,  Young W.,   Golap K.,  2007, in
  Astronomical {{Data Analysis Software}} and {{Systems XVI}}. p.~127

\bibitem[\protect\citeauthoryear{Migliari \& Fender}{Migliari \&
  Fender}{2006}]{Migliari2006}
Migliari S.,  Fender R.~P.,  2006, \mn@doi [Mon. Not. Roy. Astron. Soc.]
  {10.1111/j.1365-2966.2005.09777.x}, 366, 79

\bibitem[\protect\citeauthoryear{{Miller-Jones}, {Jonker}, {Dhawan}, {Brisken},
  {Rupen}, {Nelemans}  \& {Gallo}}{{Miller-Jones}
  et~al.}{2009}]{Miller-Jones2009}
{Miller-Jones} J.~C.~A.,  {Jonker} P.~G.,  {Dhawan} V.,  {Brisken} W.,  {Rupen}
  M.~P.,  {Nelemans} G.,   {Gallo} E.,  2009, \mn@doi [\apjl]
  {10.1088/0004-637X/706/2/L230}, \href
  {https://ui.adsabs.harvard.edu/abs/2009ApJ...706L.230M} {706, L230}

\bibitem[\protect\citeauthoryear{{Miller} et~al.,}{{Miller}
  et~al.}{2020}]{Miller2020}
{Miller} J.~M.,  et~al., 2020, \mn@doi [\apj] {10.3847/1538-4357/abbb2e}, \href
  {https://ui.adsabs.harvard.edu/abs/2020ApJ...904...49M} {904, 49}

\bibitem[\protect\citeauthoryear{Nardiello et~al.,}{Nardiello
  et~al.}{2018}]{Nardiello2018}
Nardiello D.,  et~al., 2018, \mn@doi [Monthly Notices of the Royal Astronomical
  Society] {10.1093/mnras/sty2515}, 481, 3382

\bibitem[\protect\citeauthoryear{Papitto \& Torres}{Papitto \&
  Torres}{2015}]{Papitto2015}
Papitto A.,  Torres D.~F.,  2015, arXiv:1504.05029 [astro-ph]

\bibitem[\protect\citeauthoryear{Papitto et~al.}{Papitto
  et~al.}{2013}]{Papitto2013}
Papitto A.,  et~al., 2013, \mn@doi [Nature] {10.1038/nature12470}, 501, 517

\bibitem[\protect\citeauthoryear{{Patterson} \& {Raymond}}{{Patterson} \&
  {Raymond}}{1985}]{Patterson1985}
{Patterson} J.,  {Raymond} J.~C.,  1985, \mn@doi [\apj] {10.1086/163187}, \href
  {https://ui.adsabs.harvard.edu/abs/1985ApJ...292..535P} {292, 535}

\bibitem[\protect\citeauthoryear{Pickles}{Pickles}{1998}]{Pickles1998}
Pickles A.~J.,  1998, \mn@doi [Publications of the Astronomical Society of the
  Pacific] {10.1086/316197}, 110, 863

\bibitem[\protect\citeauthoryear{{Plotkin}, {Gallo}  \& {Jonker}}{{Plotkin}
  et~al.}{2013}]{Plotkin2013}
{Plotkin} R.~M.,  {Gallo} E.,   {Jonker} P.~G.,  2013, \mn@doi [\apj]
  {10.1088/0004-637X/773/1/59}, \href
  {https://ui.adsabs.harvard.edu/abs/2013ApJ...773...59P} {773, 59}

\bibitem[\protect\citeauthoryear{Plotkin et~al.,}{Plotkin
  et~al.}{2017}]{Plotkin2017}
Plotkin R.~M.,  et~al., 2017, \mn@doi [The Astrophysical Journal]
  {10.3847/1538-4357/834/2/104}, 834, 104

\bibitem[\protect\citeauthoryear{Rea et~al.}{Rea et~al.}{2017}]{Rea2017}
Rea N.,  et~al., 2017, \mn@doi [Mon. Not. Roy. Astron. Soc.]
  {10.1093/mnras/stx1560}, 471, 2902

\bibitem[\protect\citeauthoryear{{Reynolds}, {Reis}, {Miller}, {Cackett}  \&
  {Degenaar}}{{Reynolds} et~al.}{2014}]{Reynolds2014}
{Reynolds} M.~T.,  {Reis} R.~C.,  {Miller} J.~M.,  {Cackett} E.~M.,
  {Degenaar} N.,  2014, \mn@doi [\mnras] {10.1093/mnras/stu832}, \href
  {https://ui.adsabs.harvard.edu/abs/2014MNRAS.441.3656R} {441, 3656}

\bibitem[\protect\citeauthoryear{Rivera~Sandoval et~al.,}{Rivera~Sandoval
  et~al.}{2018}]{Sandoval2018}
Rivera~Sandoval L.~E.,  et~al., 2018, \mn@doi [Monthly Notices of the Royal
  Astronomical Society] {10.1093/mnras/sty058}, 475, 4841

\bibitem[\protect\citeauthoryear{Russell et~al.,}{Russell
  et~al.}{2016}]{Russell2016}
Russell T.~D.,  et~al., 2016, \mn@doi [Monthly Notices of the Royal
  Astronomical Society] {10.1093/mnras/stw1238}, 460, 3720

\bibitem[\protect\citeauthoryear{{Shahbaz}, {Linares}, {Rodr{\'\i}guez-Gil}  \&
  {Casares}}{{Shahbaz} et~al.}{2019}]{Shahbaz2019}
{Shahbaz} T.,  {Linares} M.,  {Rodr{\'\i}guez-Gil} P.,   {Casares} J.,  2019,
  \mn@doi [\mnras] {10.1093/mnras/stz1652}, \href
  {https://ui.adsabs.harvard.edu/abs/2019MNRAS.488..198S} {488, 198}

\bibitem[\protect\citeauthoryear{Shishkovsky et~al.,}{Shishkovsky
  et~al.}{2020}]{Shishkovsky2020}
Shishkovsky L.,  et~al., 2020, \mn@doi [\apj] {10.3847/1538-4357/abb880}, 903,
  73

\bibitem[\protect\citeauthoryear{Stacey, Heinke, Cohn, Lugger  \&
  Bahramian}{Stacey et~al.}{2012}]{Stacey2012}
Stacey W.~S.,  Heinke C.~O.,  Cohn H.~N.,  Lugger P.~M.,   Bahramian A.,  2012,
  \mn@doi [The Astrophysical Journal] {10.1088/0004-637X/751/1/62}, 751, 62

\bibitem[\protect\citeauthoryear{Stappers et~al.}{Stappers
  et~al.}{2014}]{Stappers2014}
Stappers B.~W.,  et~al., 2014, \mn@doi [Astrophys. J.]
  {10.1088/0004-637X/790/1/39}, 790, 39

\bibitem[\protect\citeauthoryear{{Strader} et~al.,}{{Strader}
  et~al.}{2019}]{Strader2019}
{Strader} J.,  et~al., 2019, \mn@doi [\apj] {10.3847/1538-4357/aafbaa}, \href
  {https://ui.adsabs.harvard.edu/abs/2019ApJ...872...42S} {872, 42}

\bibitem[\protect\citeauthoryear{{Strader} et~al.,}{{Strader}
  et~al.}{2021}]{Strader2021}
{Strader} J.,  et~al., 2021, arXiv e-prints, \href
  {https://ui.adsabs.harvard.edu/abs/2021arXiv210607657S} {p. arXiv:2106.07657}

\bibitem[\protect\citeauthoryear{Swihart et~al.,}{Swihart
  et~al.}{2018}]{Swihart2018}
Swihart S.~J.,  et~al., 2018, \mn@doi [\apj] {10.3847/1538-4357/aadcab}, 866,
  83

\bibitem[\protect\citeauthoryear{{Tendulkar} et~al.,}{{Tendulkar}
  et~al.}{2014}]{Tendulkar2014}
{Tendulkar} S.~P.,  et~al., 2014, \mn@doi [\apj] {10.1088/0004-637X/791/2/77},
  \href {https://ui.adsabs.harvard.edu/abs/2014ApJ...791...77T} {791, 77}

\bibitem[\protect\citeauthoryear{{The Fermi LAT collaboration}}{{The Fermi LAT
  collaboration}}{2010}]{TheFermiLATcollaboration2010}
{The Fermi LAT collaboration} 2010, \mn@doi [Astronomy \& Astrophysics]
  {10.1051/0004-6361/201014458}, 524, A75

\bibitem[\protect\citeauthoryear{Tody}{Tody}{1986}]{Tody1986}
Tody D.,  1986, \mn@doi [Instrumentation in astronomy VI] {10.1117/12.968154},
  627, 733

\bibitem[\protect\citeauthoryear{{Tody}}{{Tody}}{1993}]{Tody1993}
{Tody} D.,  1993, in {Hanisch} R.~J.,  {Brissenden} R.~J.~V.,   {Barnes} J.,
  eds,  Astronomical Society of the Pacific Conference Series Vol. 52,
  Astronomical Data Analysis Software and Systems II. p.~173

\bibitem[\protect\citeauthoryear{Tremou et~al.,}{Tremou
  et~al.}{2018}]{Tremou2018}
Tremou E.,  et~al., 2018, \mn@doi [The Astrophysical Journal]
  {10.3847/1538-4357/aac9b9}, 862, 16

\bibitem[\protect\citeauthoryear{{Tudor} et~al.,}{{Tudor}
  et~al.}{2017}]{Tudor2017a}
{Tudor} V.,  et~al., 2017, \mn@doi [\mnras] {10.1093/mnras/stx1168}, \href
  {https://ui.adsabs.harvard.edu/abs/2017MNRAS.470..324T} {470, 324}

\bibitem[\protect\citeauthoryear{{Vilhu} \& {Walter}}{{Vilhu} \&
  {Walter}}{1987}]{Vilhu1987}
{Vilhu} O.,  {Walter} F.~M.,  1987, \mn@doi [\apj] {10.1086/165689}, \href
  {https://ui.adsabs.harvard.edu/abs/1987ApJ...321..958V} {321, 958}

\bibitem[\protect\citeauthoryear{Wijnands \& {van der Klis}}{Wijnands \& {van
  der Klis}}{1998}]{Wijnands1998a}
Wijnands R.,  {van der Klis} M.,  1998, \mn@doi [Nature] {10.1038/28557}, 394,
  344

\bibitem[\protect\citeauthoryear{Wilms, Allen  \& McCray}{Wilms
  et~al.}{2000}]{Wilms2000}
Wilms J.,  Allen A.,   McCray R.,  2000, \mn@doi [The Astrophysical Journal]
  {10.1086/317016}, 542, 914

\bibitem[\protect\citeauthoryear{Yungelson, Kuranov  \& Postnov}{Yungelson
  et~al.}{2019}]{Yungelson2019}
Yungelson L.~R.,  Kuranov A.~G.,   Postnov K.~A.,  2019, \mn@doi
  [\textbackslash mnras] {10.1093/mnras/stz467}, 485, 851

\bibitem[\protect\citeauthoryear{Zhao, Heinke, Cohn, Lugger  \& Cool}{Zhao
  et~al.}{2019}]{Zhao2019}
Zhao Y.,  Heinke C.~O.,  Cohn H.~N.,  Lugger P.~M.,   Cool A.~M.,  2019,
  \mn@doi [Monthly Notices of the Royal Astronomical Society]
  {10.1093/mnras/sty3384}, 483, 4560

\bibitem[\protect\citeauthoryear{{de Martino} et~al.,}{{de Martino}
  et~al.}{2010}]{deMartino2010}
{de Martino} D.,  et~al., 2010, \mn@doi [Astronomy and Astrophysics]
  {10.1051/0004-6361/200913802}, 515, A25

\bibitem[\protect\citeauthoryear{{de Martino} et~al.,}{{de Martino}
  et~al.}{2013}]{deMartino2013}
{de Martino} D.,  et~al., 2013, \mn@doi [Astronomy \& Astrophysics]
  {10.1051/0004-6361/201220393}, 550, A89

\makeatother
\end{thebibliography}








\bsp	
\label{lastpage}
\end{document}